\documentclass[aps, prd, showkeys,superscriptaddress, nofootinbib, floatfix]{revtex4-2}

\usepackage{amsmath}
\usepackage{graphicx}
\usepackage{dcolumn}
\usepackage{bm}

\usepackage[bookmarksnumbered, pdfpagelabels=true, plainpages=false, colorlinks=true, linkcolor=blue, citecolor=blue, urlcolor=blue]{hyperref}

\usepackage{amssymb}
\usepackage{dsfont}
\usepackage{url}
\usepackage[justification=Justified,
   format=plain]{caption}
\usepackage{subcaption}
\usepackage{physics}
\usepackage{xcolor}
\usepackage{geometry}

\begin{document}

\title{Analytical insights into the interplay of momentum, multiplicity and the speed of sound in heavy-ion collisions}

\author{Gabriel Soares Rocha}
\email{gabriel.soares.rocha@vanderbilt.edu}
\affiliation{Department of Physics and Astronomy, Vanderbilt University, Nashville TN 37240}

\author{Lorenzo Gavassino}
\email{lorenzo.gavassino@vanderbilt.edu}
\affiliation{Department of Mathematics, Vanderbilt University, Nashville TN 37240}

\author{Mayank Singh}
\email{mayank.singh@vanderbilt.edu}
\affiliation{Department of Physics and Astronomy, Vanderbilt University, Nashville TN 37240}

\author{Jean-François Paquet}
\email{jean-francois.paquet@vanderbilt.edu}
\affiliation{Department of Physics and Astronomy, Vanderbilt University, Nashville TN 37240}
\affiliation{Department of Mathematics, Vanderbilt University, Nashville TN 37240}

\begin{abstract}
We introduce a minimal model of ultracentral heavy-ion collisions to study the relation between the speed of sound of the produced plasma and the final particles' energy and multiplicity. 
We discuss how the particles'
multiplicity $N_{\textrm{tot}}$ and average energy $E_{\textrm{tot}}/N_{\textrm{tot}}$ is related to the speed of sound $c_s$ by 
$c_s^2=d \ln (E_{\textrm{tot}}/N_{\textrm{tot}})/d\ln N_{\textrm{tot}}$ if the fluid is inviscid, its speed of sound is constant and all final particles can be measured. We show that finite rapidity cuts on the particles' multiplicity $N$ and energy $E$ introduce corrections between $c_s^2$ and $d \ln (E/N)/d\ln N$ that depend on the system's lifetime.
We study analytically these deviations with the Gubser hydrodynamic solution, finding that, for ultrarelativistic bosons, they scale as the ratio of the freezeout temperature $T_{\mathrm{FO}}$ over the maximum initial temperature of the fluid $T_{0}$; the non-thermodynamic aspect of these corrections is highlighted through their dependence on the system's initial conditions.
\end{abstract}

\maketitle

\section{Introduction}

Ultrarelativistic heavy-ion collisions can produce strongly-coupled quark-gluon plasma, a state of matter where the fundamental degrees of freedom of Quantum Chromodynamics (QCD) are deconfined \cite{Harris:2024aov,Heinz:2013th,Ratti:2021ubw,Yagi:2005yb}. The plasma expands at velocities of the order of the speed of light, and its evolution can be described with relativistic hydrodynamic models~\cite{Heinz:2013th,Gale:2013da,Busza:2018rrf,DerradideSouza:2015kpt,Niida:2021wut}. Recombining into hadrons as it cools down, the plasma's properties can be studied from the resulting final particles that reach the detectors.

The primary connection point between the properties of the plasma and the equations of fluid dynamics is the equation of state. At zero or small values of baryon chemical potential $\mu_B$, the equation of state of nuclear matter has been computed \textit{ab initio} using lattice QCD simulations \cite{Bazavov:2009zn,Borsanyi:2010cj,Bazavov:2017dus,Borsanyi:2021sxv,Ratti:2018ksb}. The zero-$\mu_B$ lattice equation of state, matched at low temperature to a hadron resonance gas~\cite{Venugopalan:1992hy,Karsch:2003vd,Huovinen:2009yb,Bazavov:2009zn,Borsanyi:2010cj,Ratti:2021ubw}, is almost universally used as a pre-determined input for hydrodynamic simulations of high-energy heavy-ion collisions.\footnote{There is, however, much interest in constraining from data the equation of state at \emph{finite} baryon chemical potential. See Refs.~\cite{Sorensen:2023zkk,MUSES:2023hyz,Du:2024wjm} for recent reviews.} While there are known subtleties in matching the lattice equation of state to a hadron resonance gas~\cite{Auvinen:2020mpc,Moreland:2015dvc}, 
these issues are likely much smaller than the broader uncertainties present in heavy-ion collision modeling.
Consequently, there have been few recent attempts to extract the equation of state from heavy-ion measurements, with most efforts focusing on other properties of QGP like its viscosity.\footnote{Ref.~\cite{Auvinen:2020mpc} studied simultaneously the uncertainties in the equation of state and their effect on data-driven constraints on the QGP viscosity.}
One notable extraction of the equation of state from heavy-ion measurements is ref.~\cite{Pratt:2015zsa}; it compared heavy-ion simulations to measurements from the Relativistic Heavy-Ion Collider (RHIC) and the Large Hadron Collider (LHC) using Bayesian parameter inference, constraining simultaneously the equation of state and other model parameters, and finding results consistent with lattice calculations for the equation of state.

Building on earlier works~\cite{Blaizot:1987cc,Campanini:2011bj,Monnai:2017cbv}, it was recently proposed in refs.~\cite{Gardim:2019xjs,Gardim:2019brr} that the QGP equation of state can be measured more directly, by using the relation between the mean transverse momentum $\langle p_{T} \rangle$ and particle multiplicity $d N_{\mathrm{ch}}/d \eta$ 
from ultra-central heavy-ion collisions to extract the speed of sound $c_s$:
\begin{equation}
\label{eq:cs2-observable}
\begin{aligned}
c_{s}^{2}(T_{\mathrm{eff}}) 
=
\frac{ d\ln \langle p_{T} \rangle}{d \ln \left[ d N_{\mathrm{ch}}/d \eta\right]} , 
\end{aligned}    
\end{equation}
where $T_{\mathrm{eff}}$ is an effective temperature of an associated uniform fluid at rest possessing the same total energy and entropy as the QGP at freezeout. The argument behind eq.~\eqref{eq:cs2-observable} in refs.~\cite{Gardim:2019xjs,Gardim:2019brr} is that the speed of sound is defined as $c_{s}^{2} \equiv dP/d \varepsilon = d \ln T/d \ln s$, the effective temperature is related to the mean transverse momentum of particles, $T_{\mathrm{eff}} \approx (1/3)\langle p_{T} \rangle$ \cite{VanHove:1982vk}, and that the entropy density is related to the particle multiplicity, $s \propto N_{\mathrm{ch}}$. 

Equation~\eqref{eq:cs2-observable} has been employed by CMS collaboration \cite{CMS:2024sgx} to extract results for the speed of sound that were found to be in very good agreement with lattice QCD calculations \cite{HotQCD:2014kol}. Subsequently, ref.~\cite{Nijs:2023bzv} employed numerical simulations of heavy-ion collisions to study mechanisms other than the equation of state that can influence the ratio $d \ln \langle p_T \rangle/d \ln \left[ d N_{\mathrm{ch}}/d \eta\right]$, identifying initial state fluctuations, energy deposition ans\"atze and centrality selection as factors.
In ref.~\cite{Gardim:2024zvi}, the authors of the original proposal assessed the robustness of eq. \eqref{eq:cs2-observable} with respect to changes in the equation of state, finding support in numerical simulations for eq.~\eqref{eq:cs2-observable}.

In this work, we study mathematically and numerically the validity of eq.~\eqref{eq:cs2-observable} using inviscid relativistic fluid dynamics with a constant speed of sound. Because the speed of sound is assumed not to vary with temperature, there is no ambiguity on which value of the speed of sound should be recovered precisely, allowing for precision studies; in particular, our results are independent of the definition of the effective temperature $T_{\mathrm{eff}}$. Given that the arguments presented in refs.~\cite{Gardim:2019xjs,Gardim:2019brr} do not involve any details of the system of interest, they should also be applicable in our simplified setting. Thus, we should be able to use our simplified model to test the essential conceptual content of the prescription.

We focus on initial temperature profiles that are fixed in shape, but whose overall normalization varies, akin to ultracentral heavy-ion collisions where the size of the energy deposition is fixed but the amount of deposited energy changes.
Variations of the normalization of the initial temperature profile lead to changes in the final particles' multiplicity and average energy, which we show in Section~\ref{sec:general_derivation} to be related to the speed of sound by
\begin{equation}
\label{eq:cs2_this_work_intro}
\begin{aligned}
c_{s}^{2} 
=
\dfrac{d\ln (E_{\text{tot}}/N_{\text{tot}})}{d\ln N_{\text{tot}}}
=
\lim_{T_{\mathrm{FO}} \to 0}
\frac{ d\ln \langle p_{T} \rangle}{d \ln N_{\textrm{mr}}} ,  
\end{aligned}    
\end{equation}
where $E_{\text{tot}}$ is the \textit{total} energy contained in the fluid and $N_{\text{tot}}$ is the \textit{total} number of particles, while $N_{\mathrm{mr}} = dN/dy \vert_{y=0}$ is the particle yield at midrapidity and $\langle p_{T} \rangle$ their midrapidity average transverse momentum.
The first equality is demonstrated to be almost exact for inviscid fluids with a constant speed of sound. 
The second equality corresponds to the case where only a subset of final particles are measured --- at midrapidity for example; in that case, the speed of sound is only recovered by taking the limit of asymptotically low freezeout temperature $T_{\mathrm{FO}}$.
We corroborate this result in Section~\ref{sec:3D} with 3+1D relativistic hydrodynamic simulations, by scanning a range of rapidity cuts for the observables. 

\begin{figure}[tb]
\centering
\begin{subfigure}{0.5\textwidth}
    \includegraphics[scale=0.26]{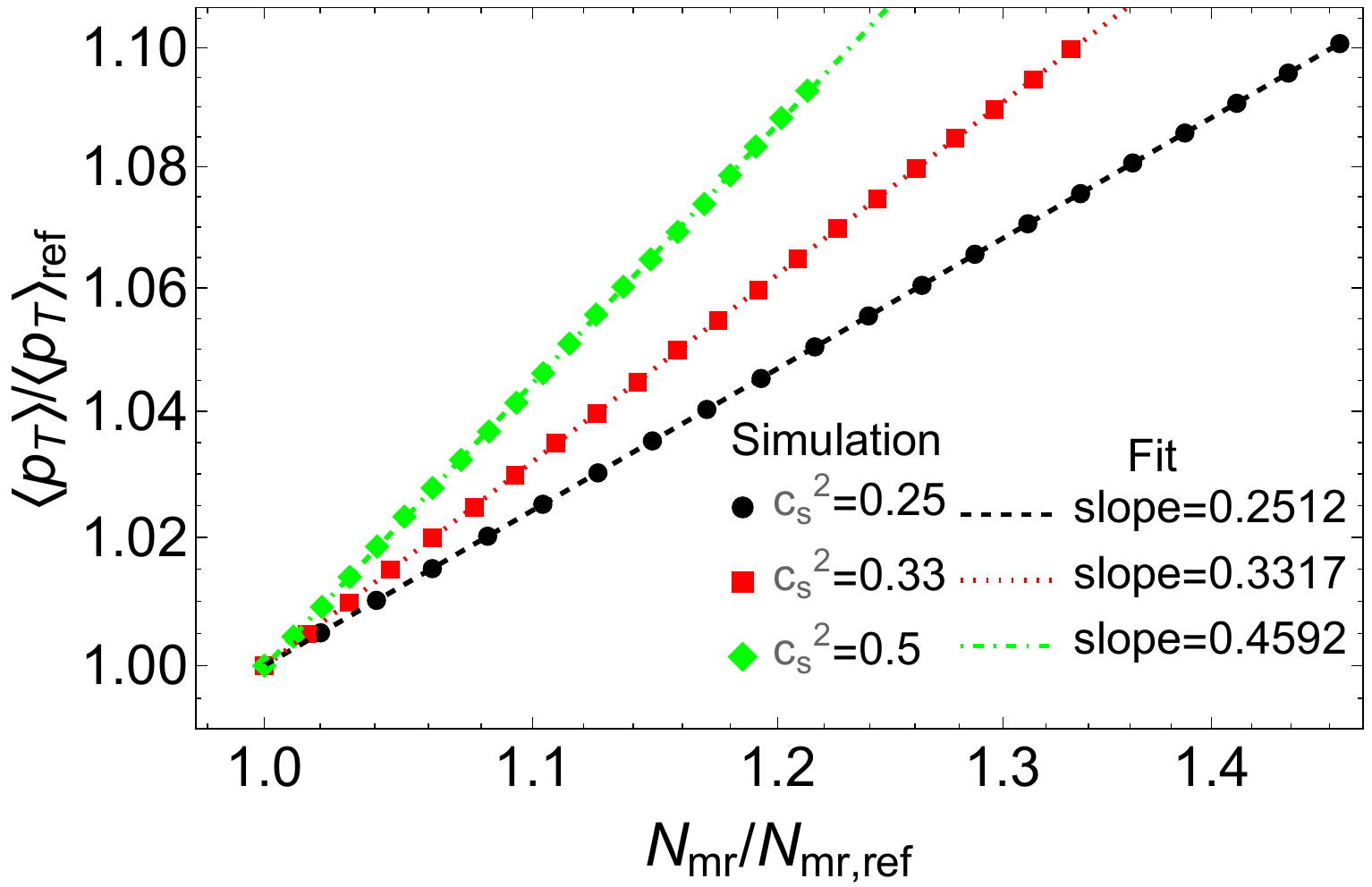}
    \caption{Particle observables at $T_{\mathrm{FO}}/T_{0} = 0.12$} 
    \label{fig:obs-tfo-60mev}
\end{subfigure}\hfil
\begin{subfigure}{0.5\textwidth}
    \includegraphics[scale=0.28]{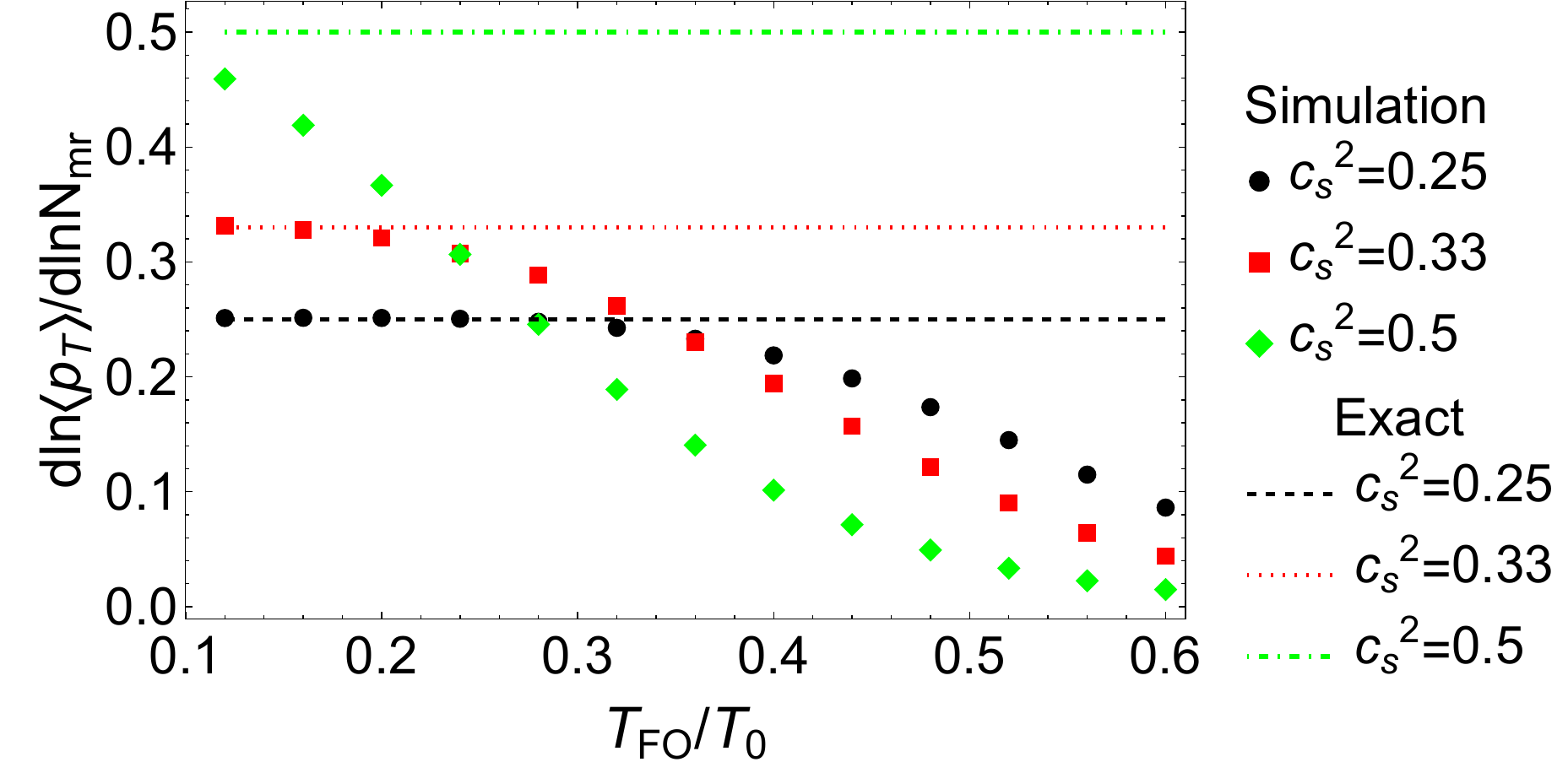}
    \caption{Speed of sound estimates $c_{s}^{2}$ observable}
    \label{fig:obs-tfo-range}
\end{subfigure}\hfil
\caption{
(a) Normalized average transverse momentum $\langle p_{T} \rangle$ as a function of normalized multiplicity $N$ (see text for normalization prescription), both at midrapidity, for three constant speeds of sound $c_{s}$, compared with the slope extracted from a linear fit of $\ln \langle p_{T} \rangle$ vs $\ln N$; 
(b) Dependence of the slope on the freezeout temperature $T_{\textrm{FO}}$. 
}
\label{fig:obs-intro}
\end{figure}

In Section~\ref{sec:cilindro}, we use a simplified 1+1D relativistic hydrodynamic model  (boost invariant and cylindrically symmetric~\cite{Baym:1983amj,Ruuskanen:1986py}) to better understand the relation between the midrapidity multiplicity and average transverse momentum and the plasma's speed of sound. Consistent with eq.~\eqref{eq:cs2_this_work_intro}, we show numerically
that this simple model can recover the speed of sound for sufficiently small freezeout temperature $T_{\mathrm{FO}}$ --- fig.~\ref{fig:obs-tfo-60mev} --- but can also exhibit a strong dependence on the plasma's lifetime, as quantified by $T_{\mathrm{FO}}$, as shown in fig.~\ref{fig:obs-tfo-range}.\footnote{We employ initial conditions with no initial transverse flow and a Gaussian initial temperature profile Gaussian: $T(\tau_{0}, r) = T_{0} e^{-r^{2}/(2\sigma^{2})}$ with $\sigma = 5$~fm.
Each point in fig.~\ref{fig:obs-tfo-60mev} represents a different globally rescaled initial condition, so that $T_{0} = \alpha T_{0}^{\star}$, with $\alpha = 1.000, 1.005, \cdots, 1.100$ and $T_{0}^{\star} = 0.500$ GeV. The model is explained in detail in Section~\ref{sec:cilindro}.}
Using the Gubser solution to inviscid relativistic hydrodynamics~\cite{Gubser:2010ui,Gubser:2010ze}, we derive analytical expressions for the multiplicity and average transverse momentum of massless (or ultrarelativistic) bosons. From these results, we quantify explicitly the dependence on $T_{\mathrm{FO}}$ of eq.~\eqref{eq:cs2_this_work_intro} in the Gubser case, finding
\begin{equation}
\frac{ d\ln \langle p_{T} \rangle}{d \ln N_{\textrm{mr}}} \approx 
\frac{1}{3} - \frac{2}{3} \frac{1}{(1 + q^{2} \tau_{0}^{2})} \left( \frac{T_{\mathrm{FO}}}{ T_{0}}\right)^{3/2}.
\end{equation}
with $T_0$ the maximum initial state temperature of the plasma and $q$ and $\tau_0$ Gubser initial condition parameters. We discuss implications for heavy-ion collisions in the summary.

{\bf Notation:} We use units such that $\hbar = c = k_{B} = 1$ and the mostly minus $(+,-,-,-)$  metric signature. For a given arbitrary four-vector $A^{\mu}$, its cylindrical coordinate components are $(A^{\mu})_{\tau,r,\phi,\eta_s}$  = $(A^{\tau}, A^{r}, A^{\phi}, A^{\eta_s})$.

\section{Total multiplicity, total energy and speed of sound in inviscid hydrodynamics}
\label{sec:general_derivation}

The derivation of eq.~\eqref{eq:cs2-observable}
 in refs.~\cite{Gardim:2019xjs,Gardim:2019brr} is based on general thermodynamic arguments.
In this section, we show that, for an inviscid fluid with constant speed of sound and specific variations of the initial conditions, it is possible to derive a nearly exact mathematical identity, whose form is close to eq.~\eqref{eq:cs2-observable}. 
We then discuss the mathematical conditions necessary for eq.~\eqref{eq:cs2-observable} 
to hold when it is not possible to measure all final state particles.

We consider a relativistic inviscid fluid without conserved charges. The fundamental quantity describing the system is the energy-momentum tensor~\cite{landau2013fluid}
\begin{equation}
\label{eq:t-munu}
\begin{aligned}
&
T^{\mu \nu} = \varepsilon u^{\mu} u^{\nu} - P \left( g^{\mu \nu} - u^{\mu} u^{\nu} \right),
\end{aligned}    
\end{equation}
where $\varepsilon$ is the energy density, $u^\mu$ the flow velocity and $P$ the pressure, which is related to the energy density $\varepsilon$ by the equation of state $P = P(\varepsilon)$. The dynamics of the system is given by the local conservation law
\begin{equation}
\begin{aligned}
&
\partial_{\mu}T^{\mu \nu} =0,
\end{aligned}    
\label{eq:dmuTmunu}
\end{equation}
which forms a closed system of partial differential equations for inviscid fluids. From eqs.~\eqref{eq:t-munu} and \eqref{eq:dmuTmunu} and the first law of thermodynamics, we also have the local conservation of the entropy current $\partial_{\mu}s^{\mu} = 0$, where $s^{\mu} = s u^{\mu}$, where $T s = \varepsilon + P$.

\subsection{A one-parameter family of solutions}
\label{sec:alphafamilona}

Let $\Psi=\{T(x^\nu),u^\mu(x^\nu) \}=\{\text{``temperature''},\text{``velocity''} \}$ be an arbitrary solution of the ideal fluid equations in 3+1 dimensions, with equation of state $P=c_s^2 \varepsilon$, where $c_s$ is the (constant) speed of sound. Let us show that the one-parameter family of spacetime profiles $\Psi(\alpha)=\{\alpha T(x^\nu),u^\mu(x^\nu) \}$, with $\alpha \in (0,+\infty)$, are also solutions to the same ideal fluid equations.

First, let us express $P$ and $\varepsilon$ as functions of $T$. Differentiating $P=c_s^2\varepsilon$, we obtain $sdT=c_s^2Tds$, where $s$ is the entropy density. This is a separable ordinary differential equation, which gives
\begin{equation}
\label{eq:s-e-p-cs2}
\begin{split}
s={}& A(1{+}c_s^{2}) \, T^{c_s^{-2}} \, ,\\
\varepsilon={}& A \, T^{1+c_s^{-2}} \, , \\
P={}& c_s^2 A \, T^{1+c_s^{-2}} \, , \\
\end{split}
\end{equation}
for some integration constant $A$. It follows that the entropy current and the stress-energy tensor are given by the following constitutive relations:
\begin{equation}
\begin{split}
s^\mu ={}& A(1{+}c_s^{2}) \, T^{c_s^{-2}} u^\mu \, , \\
T^{\mu \nu} ={}& A T^{1+c_s^{-2}} \big[ (1{+}c_s^{2})u^\mu u^\nu-c_s^2 g^{\mu \nu} \big] \, . \\
\end{split}
\end{equation}
Then, along the family $\Psi(\alpha)=\{\alpha T,u^\mu \}$, the entropy current and stress-energy tensor obey the following scaling law:
\begin{equation}\label{scalinglaw}
\begin{split}
s^\mu(\alpha)={}& \alpha^{c_s^{-2}} s^\mu(1) \, , \\
T^{\mu\nu}(\alpha)={}& \alpha^{1+c_s^{-2}} T^{\mu \nu}(1) \, . \\
\end{split}
\end{equation}
Since $\alpha$ is a constant (in spacetime), we find that $\partial_\mu s^\mu(\alpha)=\partial_\mu T^{\mu \nu}(\alpha)=0$ if and only if $\partial_\mu s^\mu(1)=\partial_\mu T^{\mu \nu}(1)=0$. Considering that the hydrodynamic equations are just the conservation laws $\partial_\mu T^{\mu \nu}=0$, we conclude that, if $\Psi=\{T,u^\nu \}$ is a solution of ideal fluid dynamics, then the whole family $\Psi(\alpha)=\{\alpha T ,u^\nu \}$ is a set of equally acceptable solutions of the same equations.

\subsection{An identity with the total energy and multiplicity}

\label{sec:cs2_tot}

Consider again the scaling law \eqref{scalinglaw}. If we integrate it along an \textit{arbitrary} Cauchy surface $\Sigma$, we obtain
\begin{equation}
\begin{split}
S_{\text{tot}}(\alpha)={}& \alpha^{c_s^{-2}} S_{\text{tot}}(1) \, , \\
E_{\text{tot}}(\alpha)={}& \alpha^{1+c_s^{-2}} E_{\text{tot}}(1) \, , \\
\end{split}
\end{equation}
where $S_{\text{tot}}$ and $E_{\text{tot}}$ are respectively the total entropy and the total energy (the latter evaluated in the laboratory frame). Thus, we immediately find that 
\begin{equation}
\begin{split}\label{etotuz}
\dfrac{d\ln S_{\text{tot}}}{d\ln\alpha} ={}& \dfrac{1}{c_s^2} \, , \\
\dfrac{d\ln E_{\text{tot}}}{d\ln\alpha} ={}& 1+ \dfrac{1}{c_s^2} \, , \\
\end{split}
\end{equation}
Note that, by Gauss' theorem, $S_{\text{tot}}$ is independent of the Cauchy surface $\Sigma$ we chose (recall that the fluid is ideal, and thus $\partial_\mu s^\mu=0$). In particular, we can take $\Sigma$ to be the freezeout surface $\Sigma_{\mathrm{FO}}$, upon which $T=T_{\mathrm{FO}}$. 
Then, considering that both the entropy density, $s$, and the particle number density, $n$, are functions only of the temperature, and thus are constant across $\Sigma_{\mathrm{FO}}$, we can take them out from the hypersurface integral (independently from the choice of equation of state). Hence, we have the following sequence of identities\footnote{It should be mentioned that, strictly speaking, the hypersurface $\Sigma_{\text{FO}}$ is not an actual Cauchy surface, because it does not cover all space, and it possesses a compact two-dimensional border. In fact, since the plasma has a finite size, $T$ tends to zero at spacelike infinity, meaning that $T$ remains below $T_{\text{FO}}$ \textit{at all times} far from the collision. Hence, one should  actually write $S_\text{tot}=S[\Sigma_\text{FO}]{+}S[\Sigma_\text{Rest}]=\int_{\Sigma_{\text{FO}}} \! \! s^\mu d\Sigma_\mu {+}\int_{\Sigma_{\text{Rest}}} \! \! s^\mu d\Sigma_\mu$ (and similarly for $N_\text{tot}$), where $\Sigma_\text{Rest}$ ``completes'' the Cauchy surface, joining the 2D-border of $\Sigma_{\text{FO}}$ to spacelike infinity. In practice, $S[\Sigma_\text{Rest}]$ and $N[\Sigma_\text{Rest}]$ are negligibly small in heavy-ion collisions, so we can say that \eqref{thebossequation} is almost exact. Note that, if $T_\text{FO}\rightarrow 0$, the identities in \eqref{thebossequation} become \textit{truly} exact, since $\Sigma_\text{Rest}\rightarrow \emptyset$.}: 
\begin{equation}\label{thebossequation}
S_{\text{tot}} = \int_{\Sigma_{\mathrm{FO}}} s(T_{\mathrm{FO}}) u^\mu d\Sigma_\mu = s(T_{\mathrm{FO}})\int_{\Sigma_{\mathrm{FO}}} u^\mu d\Sigma_\mu =\dfrac{s(T_{\mathrm{FO}})}{n(T_{\mathrm{FO}})}  \int_{\Sigma_{\mathrm{FO}}} n(T_{\mathrm{FO}}) u^\mu d\Sigma_\mu = \dfrac{s(T_{\mathrm{FO}})}{n(T_{\mathrm{FO}})} N_{\text{tot}} \, ,
\end{equation}
where $N_{\text{tot}}$ is the total particle number at freezeout, which (according to Cooper-Frye~\cite{Cooper:1974mv}) coincides with the total multiplicity that is later detected.\footnote{In this work, we do not use a cascade or transport model after freezeout and hence the chemical and kinetic freezeout coincide.}
Note that the value of the ratio $s(T_{\mathrm{FO}})/n(T_{\mathrm{FO}})$ does not depend on $\alpha$, because it is a pure function of $T_{\mathrm{FO}}$ (which is a fixed parameter of the system, independent from $\alpha$). Therefore, taking a derivative of the logarithm of the entropy eliminates the $s(T_{\mathrm{FO}})/n(T_{\mathrm{FO}})$ factor between $S_{\text{tot}}$ and $N_{\text{tot}}$, yielding
\begin{equation}\label{ntot}
    \dfrac{d\ln N_{\text{tot}}}{d\ln \alpha} = \dfrac{1}{c_s^2} \, .
\end{equation}
Thus, combining \eqref{etotuz} with \eqref{ntot}, we obtain the following identity:
\begin{equation}
\label{eq:cs2-lnE}
c_s^2 = \dfrac{d\ln (E_{\text{tot}}/N_{\text{tot}})}{d\ln N_{\text{tot}}} \, .
\end{equation}
Note that this relation holds for arbitrary flow geometry in $3+1$ dimensions: no symmetry assumptions are required. However, it must be kept in mind that the differential ``$d$'' is evaluated along the one-parameter family of states introduced in Section \ref{sec:alphafamilona}. None of the above relations hold for generic displacements.

\subsection{From total observables to midrapidity observables}

The identity \eqref{eq:cs2-lnE} holds when the total energy and multiplicity can be measured. For the same result to hold with midrapidity measurements, it requires the midrapidity multiplicity and the transverse energy to
share the same dependence on $\alpha$ as the corresponding total quantities, i.e.
\begin{equation}
    \dfrac{d}{d\alpha} \bigg( \dfrac{N_{\mathrm{mr}}}{N_{\text{tot}}} \bigg) = 0 \, , \quad \quad \quad \dfrac{d}{d\alpha} \bigg( \dfrac{N_{\mathrm{mr}}\langle p_T \rangle}{E_{\text{tot}}} \bigg) = 0  \, ,
    \label{eq:midrap_conditions}
\end{equation}
where we remark that, for massless particles, the transverse energy can be expressed as $dE/dy\vert_{y = 0} = N_{\mathrm{mr}}\langle p_T \rangle$ (see also Eq.~\eqref{eq:observables-kappa-T}).
Conditions \eqref{eq:midrap_conditions} hold for a spherically symmetric system, for example, since particles and energy are emitted equally in all directions. Hence, the ratio between the number of transverse particles and the total number of particles is dictated by symmetry, and it does not depend on $\alpha$. The same reasoning holds for the energy. Unfortunately, the general case (in the absence of symmetries) is not so straightforward. In the following, we will focus for clarity on the scaling behavior of the multiplicity, but a similar argument applies also to the energy.

The quantity $N_{\mathrm{mr}}$, which is the number of particles emitted in the transversal direction, depends on $\alpha$ in two ways. First, when we rescale the solution by $\alpha$, the particle current $J^\mu$ is rescaled by a factor $\alpha^{c_s^{-2}}$. This same effect enters also $N_{\text{tot}}$, and these $\alpha$-related factors cancel out. The second effect is that, along the solution $\Psi(\alpha)$, the freezeout surface is defined by the condition $T(\alpha)=\alpha T(1)=T_{\mathrm{FO}}$, which depends on $\alpha$. This means, when we change $\alpha$, we are sampling $J^\mu$ on different surfaces. It follows that, instead of varying $\alpha$, we may just set $\alpha=1$, and replace the derivative in $\alpha$ with a derivative in the freezeout temperature:
\begin{equation}
\dfrac{d}{d\alpha} \bigg( \dfrac{N_{\mathrm{mr}}}{N_{\text{tot}}} \bigg)  \propto \dfrac{d}{dT_{\mathrm{FO}}} \bigg( \dfrac{N_{\mathrm{mr}}}{N_{\text{tot}}} \bigg)     \, .
\end{equation}
Now, we note that the ratio $N_{\mathrm{mr}}/N_{\text{tot}}$ is only sensitive to the angular dependence of the flow, since it quantifies how many particles (with respect to the total) are traveling in the transverse direction. It is reasonable to assume that, at very late times, when the shape of the flow has relaxed to its late-time profile, $N_{\mathrm{mr}}(T_{\mathrm{FO}})/N_{\text{tot}}(T_{\mathrm{FO}})$ will approach a constant\footnote{There is a solid mathematical argument in support of this claim. Define $\varphi=N_{\mathrm{mr}}/N_{\text{tot}}$, and view it as a function of the freezeout hypersurface. At sufficiently late times, when the transversal expansion takes over the longitudinal expansion, the function $\varphi$ is monotonically non-decreasing. On the other hand, $\varphi$ is bounded above by $1$. A well-known theorem of analysis tells us that a bounded monotonic sequence must converge to a stationary value. Hence, $\varphi$ approaches a positive constant as $T_{\mathrm{FO}}$ approaches zero.}. In other words,
\begin{equation}
    \lim_{T_{\mathrm{FO}}\rightarrow 0} 
 \, \dfrac{d}{dT_{\mathrm{FO}}} \bigg( \dfrac{N_{\mathrm{mr}}}{N_{\text{tot}}} \bigg) = 0 \, .
\end{equation}
Applying the same reasoning to the transverse energy, we finally obtain the following claim:
\begin{equation}\label{eq:cs2-lnEmr}
c_s^2 = \dfrac{d\ln (E_{\text{tot}}/N_{\text{tot}})}{d\ln N_{\text{tot}}} = \lim_{T_{\mathrm{FO}}\rightarrow 0}  \dfrac{d\ln \langle p_T \rangle}{d\ln N_{\mathrm{mr}}} \, .
\end{equation}
In the next sections, we shall confirm the ideas outlined in the present section employing a three-dimensional hydrodynamic model of heavy-ion collisions and further explore the nonzero $T_{\mathrm{FO}}$ effects with $1+1$D models with cylindrical symmetry.  

\section{Numerical results in a simplified 3+1D model of ultrarelativistic heavy-ion collisions}
\label{sec:3D}

The equations of inviscid relativistic hydrodynamics can be written as
\begin{subequations}
\label{eq:eoms-hydro}
\begin{align}
&
\label{eq:eoms-hydro-e}
u^\mu \partial_\mu \varepsilon + (\varepsilon + P) \partial_\mu u^\mu = 0,
\\
&
\label{eq:eoms-hydro-u}
(\varepsilon + P) u^\nu \partial_\nu u^{\mu} = (g^{\mu \nu}-u^\mu u^\nu)\partial_\nu P,
\end{align}    
\end{subequations}
by projecting eq.~\eqref{eq:dmuTmunu} along the flow velocity $u^{\mu}$ (first equation) and orthogonal to it (second equation).  We solve these equations numerically with MUSIC \cite{Schenke:2010nt,Ryu:2015vwa,Paquet:2015lta,Ryu:2017qzn}. The equation of state is again a constant speed of sound one: $P = c_s^2 \varepsilon$. We use eq.~\eqref{eq:eoms-hydro} in Milne coordinates $\tau = \sqrt{t^{2} - z^{2}}$, the hyperbolic time, and $\eta_{s} = \tanh^{-1}(z/t)$ the \textit{spacetime} rapidity. 

We initialize hydrodynamics at $\tau = 0.4$ fm. We use the optical Glauber model to collide two lead nuclei with zero impact parameter to generate the initial energy density profile on the transverse plane. The energy densities are then multiplied with a longitudinal profile function to obtain 3D initial conditions.
We use the longitudinal profile from ref.~\cite{Hirano:2001eu}, which has a flat region near midrapidity that tapers off following Gaussian tails. In this work, the flat region extends between $\pm$ 4 units in spatial rapidity $\eta_s$. The width of the Gaussian tail is 0.6.  
The initial flow velocities in the transverse plane and in the $\eta_s$ direction are set to zero.

We convert the fluid's energy-momentum tensor $T^{\mu \nu}$ to particle degrees of freedom on a constant temperature hypersurface $\Sigma_{\mathrm{FO}}$ at temperature $T = T_{\mathrm{FO}}$
using with the Cooper-Frye formula \cite{Cooper:1974mv,Cooper:1974qi}
\begin{equation}
\label{eq:cooper-frye}
\begin{aligned}
E_{p}\frac{d N}{d^{3}p} = \frac{g}{(2 \pi)^{3}} \int_{\Sigma_{\mathrm{FO}}} d \Sigma_{\mu} p^{\mu} f(x,p),      
\end{aligned}    
\end{equation}
where $g$ is the degeneracy factor, $d \Sigma_{\mu}$ is the surface element of the constant temperature hypersurface and $f(x,p)$ is the phase-space distribution of the particles. For simplicity, we consider that the particles generated are of a single species of massless bosons:
\begin{equation}
\begin{aligned}
f(x,p) = f_{0}(x,p) =  \frac{1}{ e^{u \cdot p/T} - 1} .  
\end{aligned}    
\end{equation}
We set the degeneracy factor to $g=1$. 

\begin{figure}[h]
\centering
\begin{subfigure}{0.5\textwidth}
    \includegraphics[scale=0.55]{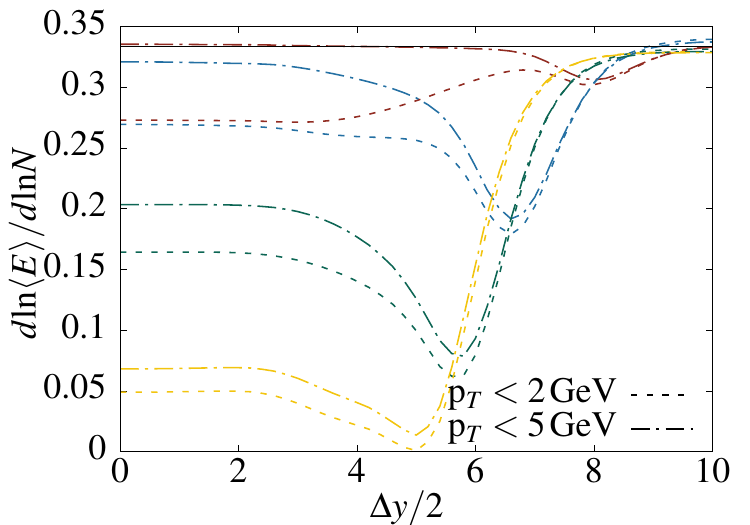}
    \caption{Slope observable as a function of rapidity \newline for $c_s^2 = 1/3$.}
    \label{fig:cs2_1b3}
\end{subfigure}\hfil
\begin{subfigure}{0.5\textwidth}
    \includegraphics[scale=0.55]{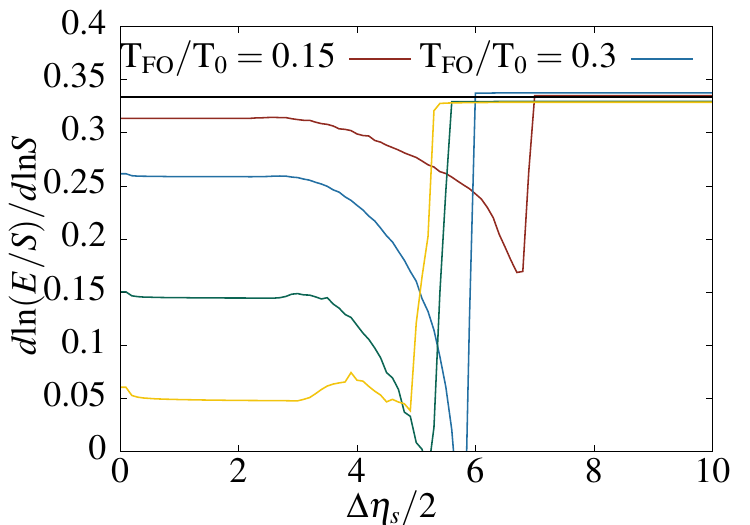}
    \caption{Slope observable from energy and entropy flux \newline for $c_s^2 = 1/3$.}
    \label{fig:surface_cs2_1b3.pdf}
\end{subfigure}\hfil
\begin{subfigure}{0.5\textwidth}
    \includegraphics[scale=0.55]{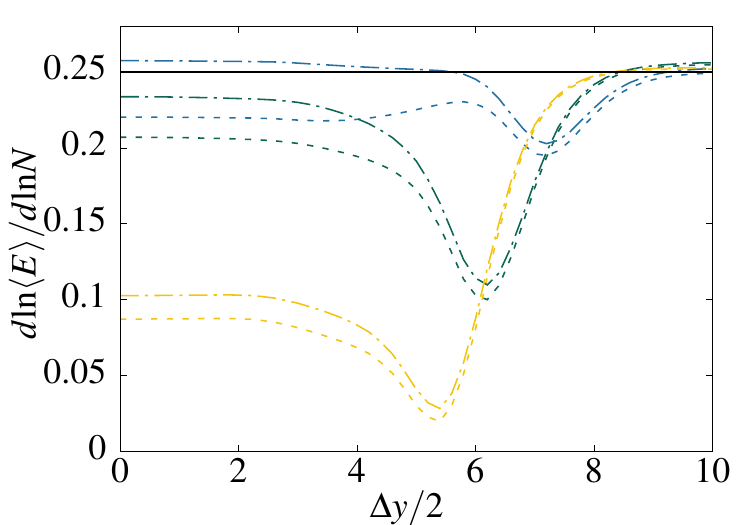}
    \caption{Slope observable as a function of rapidity \newline for $c_s^2 = 1/4$.}
    \label{fig:cs2_1b4}
\end{subfigure}\hfil
\begin{subfigure}{0.5\textwidth}
    \includegraphics[scale=0.55]{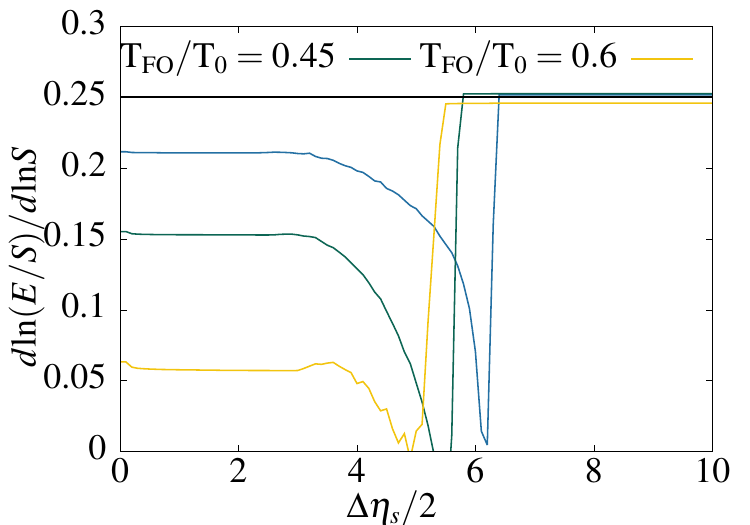}
    \caption{Slope observable from energy and entropy flux \newline for $c_s^2 = 1/4$.}
    \label{fig:surface_cs2_1b4.pdf}
\end{subfigure}\hfil
\caption{Slope observable as a function of size of rapidity window for different freezeout temperatures and $p_T$ cuts for $c_s^2 = 1/3$ (a) and $c_s^2 = 1/4$ (c). Slope estimation from energy and entropy flux through different isothermal hypersurfaces as a function of spatial rapidity window for $c_s^2 = 1/3$ (b) and  $c_s^2 = 1/4$ (d).}
\label{fig:numerical3D}
\end{figure}

We calculate the slope observable $d\ln{(E_{\rm tot}/N_{\rm tot})}/d\ln{N_{\rm tot}}$ from the particle distribution obtained after the Cooper-Frye procedure. This is done for two different equations of state, both with constant speeds of sound: $c_s^2=1/3$ and $1/4$. We evaluate the slope observable at rapidity windows of size $\Delta y$ around midrapidity. Figures~\ref{fig:cs2_1b3} and \ref{fig:cs2_1b4} show this ratio for the two different $c_s^2$. The calculation is done for two different $p_T$ cuts: we include all particles with $p_T < 2$ GeV or with $p_T < 5$ GeV. We see that for the largest rapidity window ($\Delta y/2=10$), when effectively all the particles are included, the true speed of sound is recovered. This statement is true regardless of the specific freezeout temperature chosen. This is the manifestation of the (almost) exact identity in eq.~\eqref{eq:cs2-lnE}. The same identity can be verified in figs.~\ref{fig:surface_cs2_1b3.pdf} and \ref{fig:surface_cs2_1b4.pdf}, where the ratio of total energy per unit entropy escaping the freezeout hypersurface to the total entropy is given as a function of the size of \emph{spatial} rapidity window $\Delta \eta_s/2$ around midrapidity. As we go to large enough spatial rapidity windows, all the energy and entropy is included and the true speed of sound is recovered.

The challenging issue is to recover the true speed of sound from the observables at midrapidity. From fig.~\ref{fig:numerical3D}, we see that we get closer to the true $c_s^2$ as we go to lower and lower freezeout temperatures, which is consistent with eq.~\eqref{eq:cs2-lnEmr}. 
At higher freezeout temperatures, the flow profile has not yet reached its late time limit, and the slope observable does not recover the correct $c_s^2$. The situation becomes worse when we open the rapidity window and take in more but not all particles from forward and backward rapidities. The initial temperatures are lower near the tail of the 3D distribution, causing them to freezeout even earlier, when they are farther from the late time flow profile. As the energy term is dominated by particles at large rapidities, the $c_s^2$ estimation is further away from its true value. Only when all the particles are included is the extraction of the speed of sound correct, irrespective of the specific freezeout temperature.

Similarly, $p_T$ cuts affect the quality of $c_s^2$ extraction at midrapidity, shown in figs.~\ref{fig:cs2_1b3} and \ref{fig:cs2_1b4}, as also pointed out in Ref.~\cite{Gardim:2024zvi}. Neglecting high-$p_T$ particles leads to an underestimate of the mean $p_T$ and hence underpredicts the speed of sound. In a typical ultracentral collision at RHIC or the LHC, $T_{\mathrm{FO}}\approx 100$--$150$~MeV, and $T_{0}\approx 400$--$600$~MeV, implying $T_{\mathrm{FO}}/T_{0} \approx 0.15$--$0.4$; 
the results with $T_{\mathrm{FO}}/T_{0}\geq 0.3$ depicted in Fig.~\ref{fig:numerical3D} show appreciable deviations between $c_{s}^{2}$ and its estimator.

\section{Numerical and analytical results in a simplified 1+1D model of ultrarelativistic heavy-ion collisions}
\label{sec:cilindro}

If we focus on midrapidity measurements, we can approximate the fluid as invariant under boost along the collision axis of the nuclei, simplifying the equations of inviscid hydrodynamics (eq.~\eqref{eq:eoms-hydro}). Since we are interested in ultracentral collisions, we make the further approximation that the fluid is radially symmetric in the plane transverse to the collision axis. With these two symmetries, the fluid four velocity can be parametrized as $(u^{\mu})_{\tau,r,\phi,\eta_{s}} = \left( \cosh \kappa , \sinh \kappa, 0, 0\right)$. In this setting, the two remaining independent hydrodynamic equations of motion can be cast as
\begin{subequations}
\begin{align}
&
\partial_{\tau} \ln T + \tanh \kappa \ \partial_{r} \ln T
= 
-c_s^2
\left(
\frac{1}{\tau}
+
\frac{\tanh \kappa}{r}
+
\tanh \kappa \ 
\partial_{\tau}\kappa 
+
\partial_{r}\kappa 
\right),
\\
&
\partial_{\tau} \kappa 
+
\tanh \kappa \ \partial_{r} \kappa
= - \frac{\partial_{r} \ln T}{\cosh^{2} \kappa}
+
\tanh \kappa \ c_s^2 \left(
\frac{1}{\tau}
+
\frac{\tanh \kappa}{r}
+
\tanh \kappa \ 
\partial_{\tau}\kappa 
+
\partial_{r}\kappa 
\right),
\end{align}    
\label{eq:cyl_hydro}
\end{subequations}
which can also be expressed in the simpler form \cite{Baym:1983amj,Vogt:2007zz}
\begin{subequations}
\begin{align}
&
\partial_{\tau}(T \sinh \kappa)
+
\partial_{r}(T \cosh \kappa) = 0,
\\
&
\partial_{\tau} (r \tau T^{1/c_s^{2}} \cosh \kappa) + \partial_{r}(r \tau T^{1/c_s^{2}} \sinh \kappa) = 0,
\end{align}    
\label{eq:cyl_hydro_simple}
\end{subequations}
where we have employed the fact that for fluids with constant speed of sound, $s = A(1{+}c_s^{2}) \, T^{1/c_s^{2}}$ (see eq.~\eqref{eq:s-e-p-cs2}). Then, the numerical solutions of the above equations for the temperature and flow profiles,  $T(\tau,r)$ and $\kappa(\tau,r)$, respectively, can be found. 

The Cooper-Frye particlization formula~\eqref{eq:cooper-frye} can also be simplified in this symmetric setting
\cite{Ruuskanen:1986py}:
\begin{equation}
\label{eq:yield-cylindr}
\begin{aligned}
E_{p}\frac{d N}{d^{3}p} 
=
\frac{dN}{dy d^{2}p_{T}}
=
 \frac{g}{(2 \pi )^{3}}
 \int_{\Gamma} \int_{0}^{2 \pi} d\phi \int_{\mathds{R}} d \eta_{s}    \frac{\tau  r [m_{T} \cosh(\eta_{s} - y) dr - p_{T} \cos \phi d \tau]}
 {\exp [(m_{T} \cosh \kappa \cosh(\eta_{s} -y) - p_{T} \sinh \kappa \cos \phi)/T] - 1},
\end{aligned}    
\end{equation}
where $\phi$ is the angle in the transverse plane, $y$ is the particle's momentum rapidity,  and $m_{T} = \sqrt{p^{2}_{T} + m^{2}}$ is the transverse mass for a particle of mass $m$, which reduces to $m_{T}=p_{T}$ for the massless particles we study in this work.
The integral $\int_{\Gamma}$ is a line integral on a curve $\Gamma$ in the $(\tau,r)$-plane defined by $T(\tau,r) = T_{\mathrm{FO}}$; the curve $\Gamma$  is the non-trivial part of the freezeout hypersurface, which must be determined from the solution of the hydrodynamic equations of motion. From the particle momentum distribution, the mean number of particles $dN/dy$ and the mean particle momentum $\langle p_{T} \rangle$ at a given particle momentum rapidity $y$, are
\begin{equation}
\begin{aligned}
\frac{dN}{dy}
&\equiv
\int d^{2}p_{T}
\frac{dN}{dy d^{2}p_{T}}, \\
\frac{dN}{dy} \langle p_{T} \rangle
&\equiv
\int d^{2}p_{T} p_{T}
\frac{dN}{dy d^{2}p_{T}}.
\end{aligned}    
\end{equation}

As shown in Appendix \ref{apn:details-part}, for massless particles produced at midrapidity, the formula for the multiplicity simplifies to
\begin{align}
&
\label{eq:observables-kappa-T-Nmr}
N_{\mathrm{mr}}
=
\left.
\frac{dN}{dy}
\right\vert_{y = 0}
=
\frac{2 g T^{3}_{\mathrm{FO}}}{\pi} \zeta(3) 
\int_{\Gamma} \ \tau r (dr \cosh \kappa - d\tau \sinh \kappa),
\end{align}
while the transverse energy is given by
\begin{subequations}
\label{eq:observables-kappa-T}
\begin{align}
&
\label{eq:observables-kappa-<pt>}
N_{\mathrm{mr}}
\langle p_{T} \rangle
=
\left.
\frac{dE}{dy}
\right\vert_{y = 0}
=
\frac{g T^{4}_{\mathrm{FO}}}{( 2 \pi )^{3}} \frac{4 \pi^{6} }{45} 
\int_{\Gamma} \ \tau r \left[F(\kappa) \cosh \kappa \ dr 
-
 G(\kappa) \sinh \kappa \ d\tau \right],
\\
&
\label{eq:observables-kappa-T-F}
F(\kappa) = (4 \cosh (2 \kappa)+3) \text{sech}^2\kappa E\left(-\sinh ^2\kappa\right)-4 K\left(-\sinh ^2\kappa\right) , \\
&
\label{eq:observables-kappa-G}
G(\kappa) =  \csch^2\kappa  \left[ 
(1-2 \cosh (2 \kappa)) K\left(-\sinh ^2\kappa\right)+(4 \cosh (2 \kappa)-3) E\left(-\sinh ^2\kappa\right)
\right]
\end{align}    
\end{subequations}
where $E(x) = \int_{0}^{\pi/2} d\phi (1 - x \sin^{2}\phi)^{1/2} $ is the complete elliptic integral, $K(x) =\int_{0}^{\pi/2} d\phi (1 - x \sin^{2}\phi)^{-1/2}$ is the elliptic integral of first kind, and $\zeta(s) = [1/\Gamma(s)] \int_{0}^{\infty} \mathrm{d}x \ x^{s-1}/(e^{x}-1)$ is the Riemann-zeta function and $\Gamma(z) = \int_{0}^{\infty} dt \ t^{z-1} e^{-t}$ is the Euler gamma function \cite{gradshteyn2014table}. These formulas are used in the following sections to compute the observable \eqref{eq:cs2-observable}, given solutions to inviscid hydrodynamic equations of motion, i.e.~solutions for $\kappa(\tau, r)$ and $T(\tau, r)$.

\subsection{Gaussian initial conditions}
\label{sec:cyl-sym-flow}

In order to compute the midrapidity particle yield $N_{\mathrm{mr}}$ and the mean transverse momentum $\langle p_{T} \rangle$, we employ the following initial profiles for the temperature,  
\begin{equation}
\label{eq:temp-profile}
\begin{aligned}
&
T(\tau_{0}, r) = T_{0} e^{-r^{2}/(2\sigma^{2})}.
\end{aligned}    
\end{equation}
We use $\sigma = 5$ fm as reference, based on the approximate radius of large atoms like lead and gold. We set the initial transverse velocity $u^r=\sinh(\kappa)$ to a small value ($10^{-5}$) for numerical stability, effectively equivalent to using no initial transverse flow velocity. We solve the equations in Mathematica.

\begin{figure}[ht]
\centering
\begin{subfigure}{0.5\textwidth}
    \includegraphics[scale=0.29]{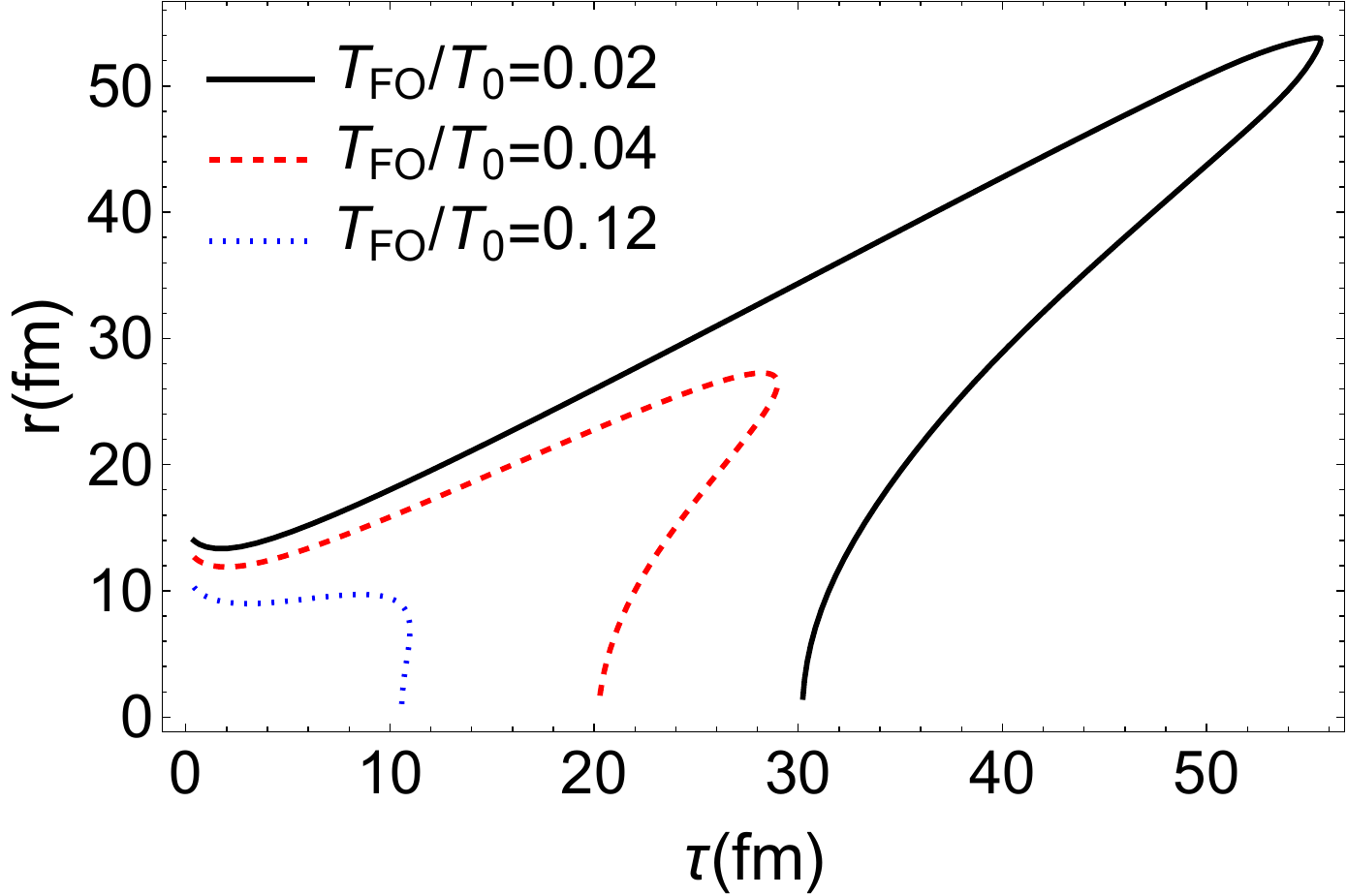}
    \caption{Low $T_{\mathrm{FO}}$ profiles $\sigma = 5.0$ fm $c_{s}^{2} = 1/3$}
    \label{fig:Tfo-profiles}
\end{subfigure}\hfil
\begin{subfigure}{0.5\textwidth}
    \includegraphics[scale=0.31]{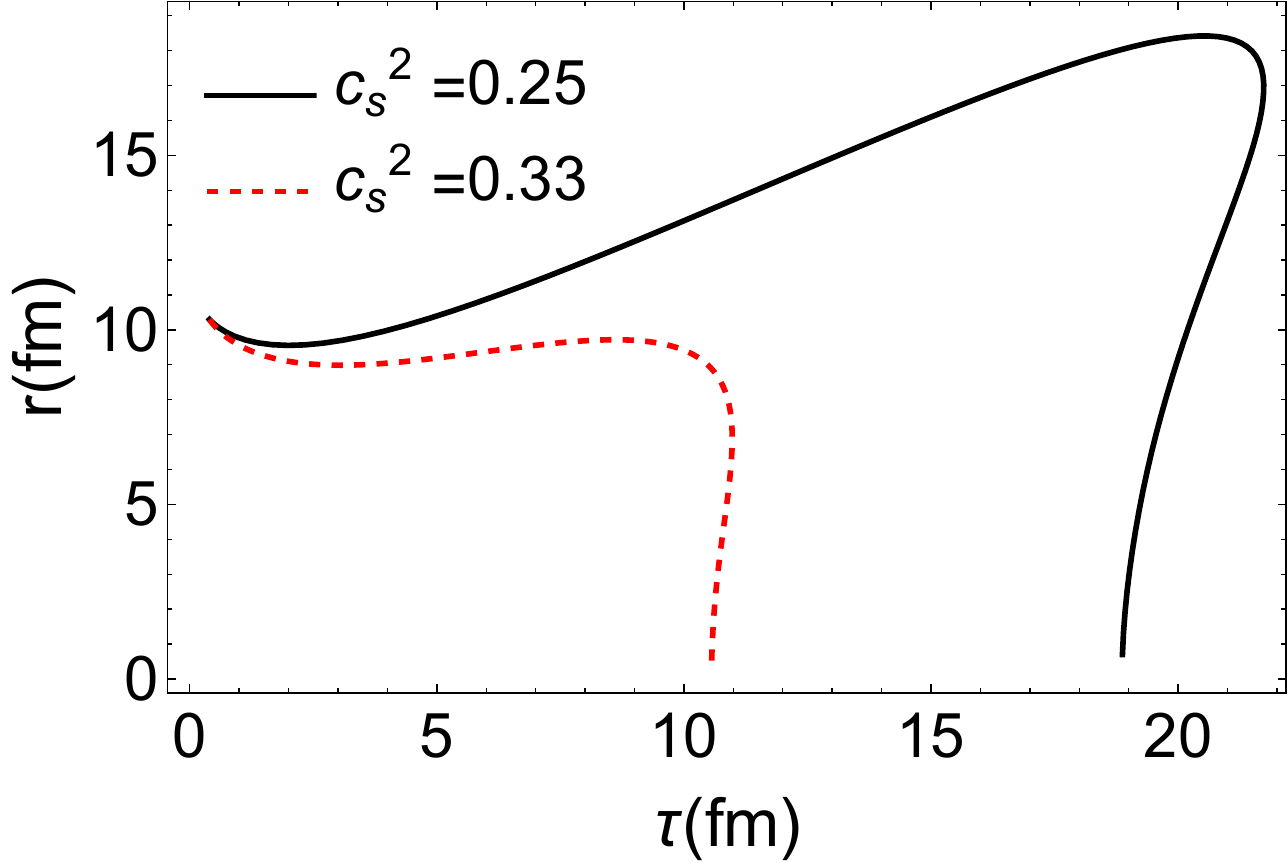}
    \caption{$T_{\mathrm{FO}}/T_{0}=0.12$ profiles for different $c_{s}^{2}$}
    \label{fig:lnN-vs-ln<pT>-cs2-cte-5}
\end{subfigure}\hfil
\caption{Temperature profiles for cylindrically symmetric, longitudinally boost-invariant systems. All results employ $\tau_{0} = 0.4$ fm $T_{0} = 0.5$ GeV.}
\label{fig:lnN-vs-ln<pT>-cs2-cte}
\end{figure}

In Fig.~\ref{fig:Tfo-profiles}, we first show the hypersurface curve $\Gamma$ from eqs.~\eqref{eq:yield-cylindr}, \eqref{eq:observables-kappa-T-Nmr} and \eqref{eq:observables-kappa-<pt>}.  We see that, the lower the temperature, the more sharply peaked the curve is around $r = \tau$ when $c_s^2=1/3$. It is from this region that most of the contribution to the integrals $N_{\mathrm{mr}}$ and $\langle p_{T} \rangle$ originate. This will become clearer when we discuss Gubser flow in the next section. Evidently, for a fixed $T_0$ and $T_{\textrm{FO}}$, the freezeout hypersurface changes significantly with $c_s^2$,
with smaller $c_{s}^{2}$ having freezeout hypersurface that spreads out further away from the origin (the propagation of the information that the system is expanding takes longer to propagate).
Measurements of the speed of sound at midrapidity are thus less accurate for
 for a fixed $T_0$ and $T_{\textrm{FO}}$, as it can be also remarked in fig.~\ref{fig:obs-intro}. The spatial spread-out of the hypersurface is confirmed in fig.~\ref{fig:lnN-vs-ln<pT>-cs2-cte-5}, where we see a significant increase in the typical distance of the freezeout hypersurface when decreasing $c_{s}^{2}$ from $1/3$ to $1/4$.

\begin{figure}[!htb]
\centering
\begin{subfigure}{0.5\textwidth}
    \includegraphics[scale=0.30]{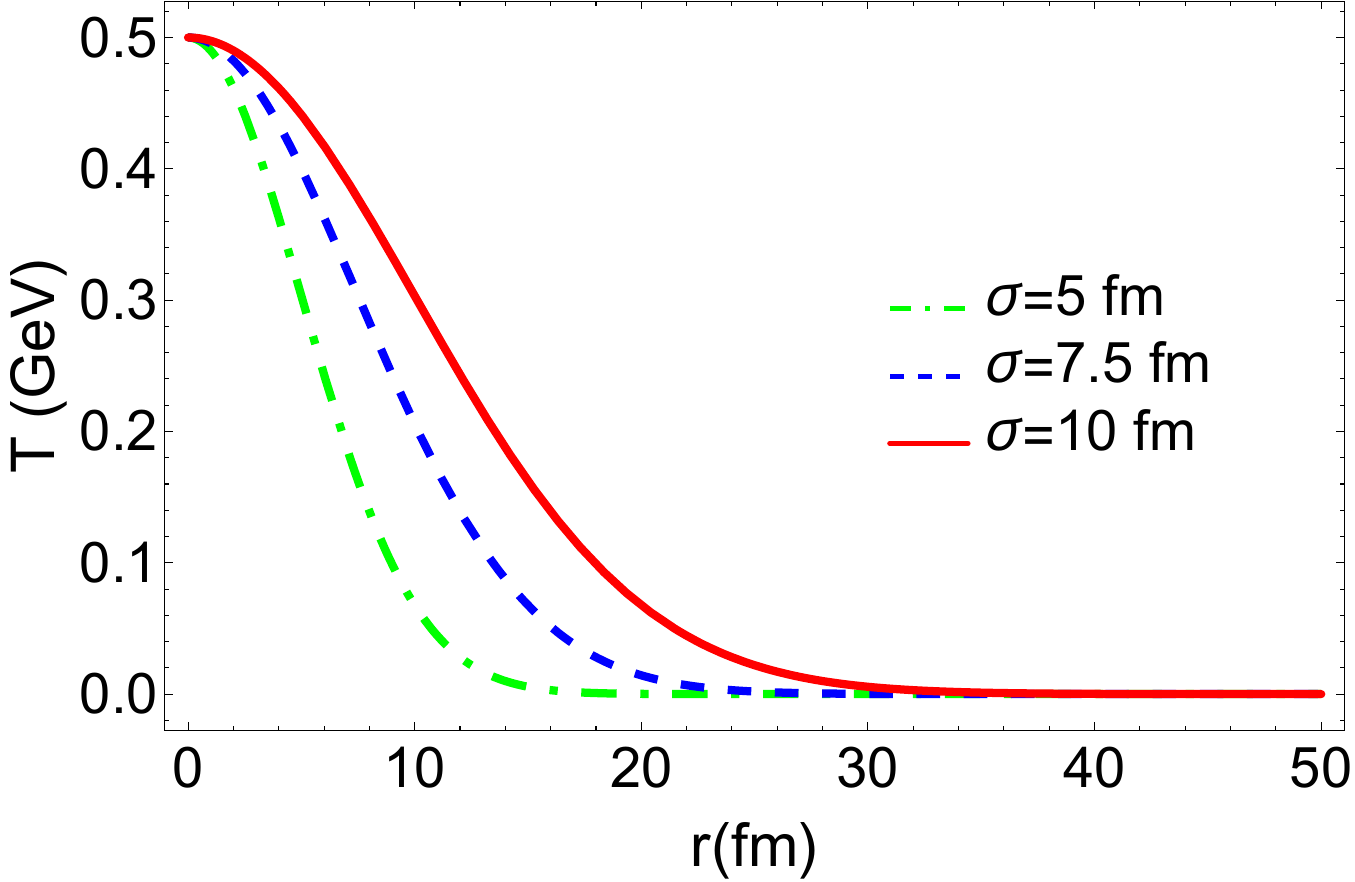}
    \caption{Initial temperature profile}
    \label{fig:T0-profiles}
\end{subfigure}\hfil
\\
\begin{subfigure}{0.5\textwidth}
    \includegraphics[scale=0.325]{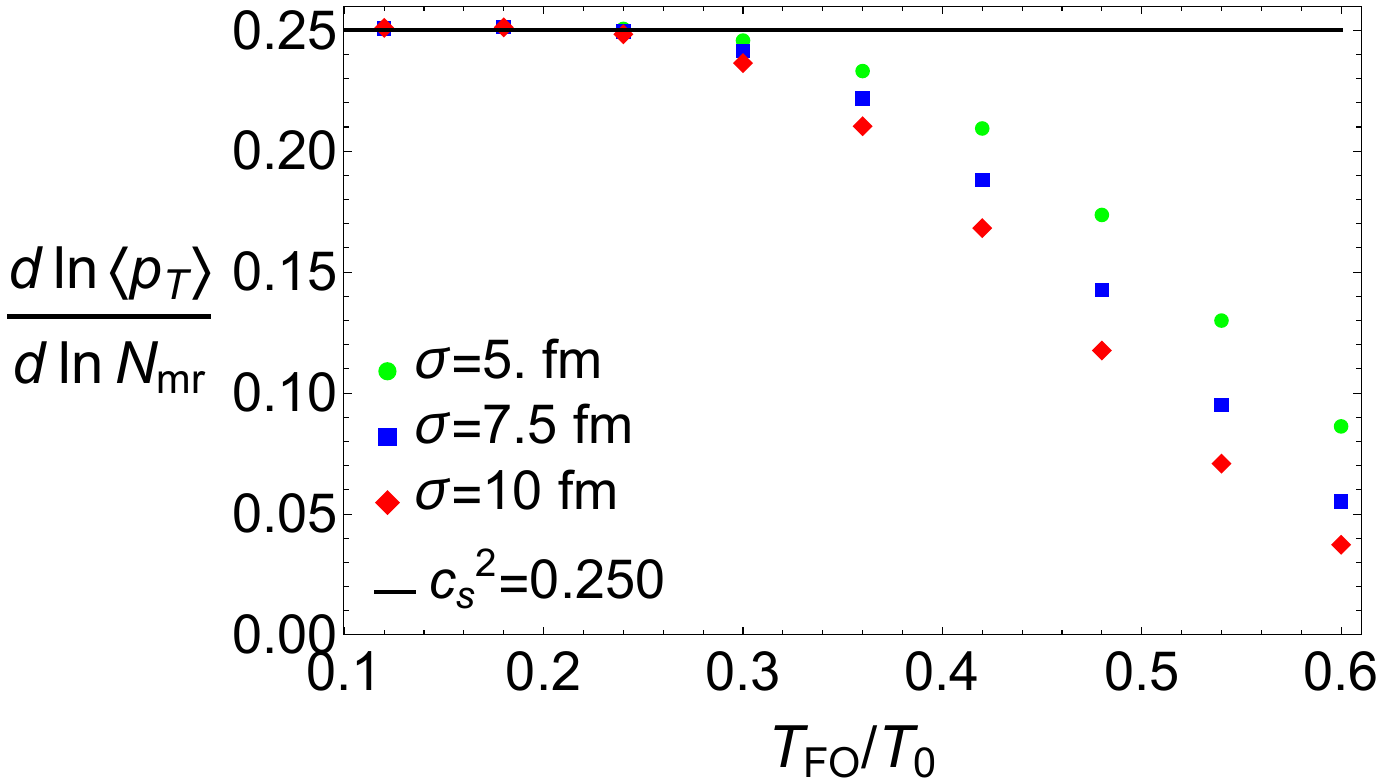}
    \caption{$c_{s}^{2} = 1/4$ }
    \label{fig:lnN-vs-ln<pT>-cs2-cte-3}
\end{subfigure}\hfil
\begin{subfigure}{0.5\textwidth}
    \includegraphics[scale=0.325]{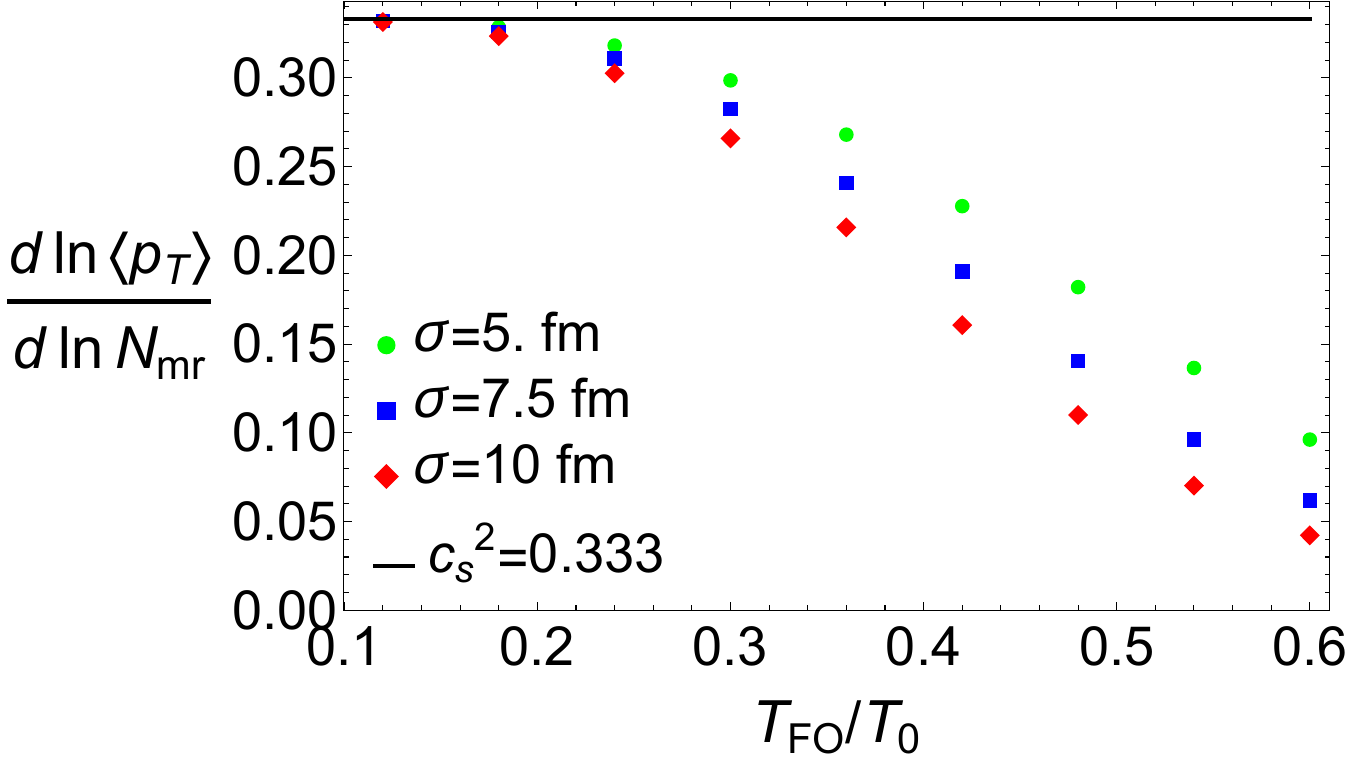}
    \caption{$c_{s}^{2} = 1/3$}
    \label{fig:lnN-vs-ln<pT>-cs2-cte-4}
\end{subfigure}\hfil
\caption{Slope observable as a function of $T_{\mathrm{FO}}$ for cylindrically symmetric, longitudinally boost-invariant systems for different initial profile width $\sigma$. All results employ $\tau_{0} = 0.4$ fm $T_{0} = 0.5$ GeV.}
\label{fig:lnN-vs-ln<pT>-cs2-cte2}
\end{figure}

Now we assess the effect of the width of initial temperature profile on the slope observable. To that end, we employ eq.~\eqref{eq:temp-profile} with $\sigma = 5,7.5,10$ fm, which are displayed in fig.~\ref{fig:T0-profiles}. Similarly to the results of fig.~\ref{fig:obs-tfo-60mev}, we employ fits to estimate $d \ln \langle p_{T} \rangle/d \ln N_{\mathrm{mr}}$ using globally-rescaled initial temperature profile, so that $T_{0} = \alpha T_{0}^{\star}$, with $\alpha = 1.000, 1.005, \cdots, 1.100$ and $T_{0}^{\star} = 0.500$ GeV. Figures \ref{fig:lnN-vs-ln<pT>-cs2-cte-3} and \ref{fig:lnN-vs-ln<pT>-cs2-cte-4}, portray results for $c_{s}^{2} = 1/4$ and $1/3$, respectively. We see once again that the slope estimates are in general more accurate the smaller the value of $T_{\mathrm{FO}}/T_{0}$ is. Moreover, in both panels we see that the agreement is also improved the smaller the value of $\sigma$ is, i.e., the more sharply peaked the initial temperature profile is.


\subsection{Gubser solution}

References \cite{Gubser:2010ze,Gubser:2010ui} showed that exact solutions of the relativistic hydrodynamics equations can be found for a conformal system with longitudinal boost invariance and radial transverse symmetry. The Gubser solutions are most conveniently expressed in a special set of curvilinear coordinates:
\begin{equation}
\begin{aligned}
&
\sinh \rho =  - \frac{1-q^{2} \tau^{2} + q^{2} r^{2}}{2 q \tau}, \
\tan \theta_{g} = \frac{2qr}{1 + q^{2} \tau^{2} - q^{2} r^{2}},
\end{aligned}
\end{equation}
where $\tau$ and $\eta_{s}$ are the Milne coordinates defined in Section~\ref{sec:3D}.
The Weyl-rescaled Minkowski line element in these coordinates reads $d\hat{s}^{2} 
=
ds^{2}/\tau^{2}
=
\hat{g}_{\mu \nu} d\hat{x}^{\mu} d\hat{x}^{\nu}
=
d \rho^{2} 
-
\cosh^{2} \rho \ d \theta^{2}_{g} 
- 
\cosh^{2} \rho \ \sin^{2} \theta d\phi^{2} 
- 
d \eta^{2}_{s}$ . With this rescaling, the dynamical variables also become rescaled, thus we have $\hat{T} = \tau T(\tau, r)$, $\hat{\varepsilon}_{0} = \tau^{4} \varepsilon_{0} (\tau, r)$, and $(u_{\nu})_{\tau, r, \phi, \eta_{s}} = \tau (u_{\nu})_{\rho, \theta_{g}, \phi, \eta} \partial \hat{x}^{\nu}/\partial x^{\mu}$. 
The main advantage of highly symmetric configurations such as the Gubser one is that $(u^{\mu})_{\rho, \theta_{g}, \phi, \eta} = (1,0,0,0)$ is a solution to the fluid conservation laws \eqref{eq:eoms-hydro} (or more specifically of eqs.~\eqref{eq:cyl_hydro} or  \eqref{eq:cyl_hydro_simple}).
The inviscid Gubser solution takes the form
\begin{equation}
\begin{aligned}
\label{eq:hydro-EoM-eps}
 \partial_{\rho}\Hat{\varepsilon}_{0} + \frac{8}{3} \Hat{\varepsilon}_{0} \tanh \rho &= 0,
\end{aligned}
\end{equation}
The equation above is readily solved by $\Hat{\varepsilon}_{0}(\rho) =
\Hat{\varepsilon}_{0}(\rho_{0}) \left(\cosh\rho_{0}/\cosh \rho\right)^{8/3}$. Transforming back to Milne coordinates, we have that the temperature and flow profiles are given, respectively by \cite{Gubser:2010ui,Gubser:2010ze,Denicol:2021}
\begin{subequations}
\label{eq:gubser-all}
\begin{align}
\label{eq:T_Gubser}
T(\tau,r)&=\frac{\hat{T}_0 (2 q \tau)^{2/3}}{\tau \left(1+2 q^2 \left(\tau^2+r^2\right)+q^4 \left(\tau^2-r^2\right)^2\right)^{1/3}},
\\
\label{eq:flow_Gubser}
\frac{u^r}{u^\tau}&=v^r(\tau,r) = \tanh \kappa(\tau,r)=\frac{2 q^2 r \tau}{1+q^2 \left(\tau^2+r^2\right)}, 
\end{align}
\end{subequations}
where $\hat{T}_0 = \tau_{0} T_{0} [(1 + q^{2} \tau_{0}^{2})/(2 q \tau_{0})]^{2/3}$. These profiles can be employed in eqs.~\eqref{eq:observables-kappa-T-Nmr} and \eqref{eq:observables-kappa-T} to compute the midrapidity particle yield $N_{\mathrm{mr}}$ and the mean transverse momentum $\langle p_{T} \rangle$ of massless bosons and perform speed of sound estimates.

As discussed in the previous sections, the slope estimate captures the value of $c_{s}^{2}$ better if we take low values of the freezeout temperature $T_{\mathrm{FO}}$. In this regime, for Gubser flow, the constant temperature curves in the $\tau$--$r$ plane become highly peaked around the line $\tau = r$, and the form of the freezeout hypersurface can be approximated by a simple function. For sufficiently low temperatures, the multiplicity and mean transverse momentum integrands in eq.~\eqref{eq:observables-kappa-T-Nmr} and eq.~\eqref{eq:observables-kappa-<pt>} are dominated by the contributions stemming from the $\tau = r$ region, which we show in Appendix \ref{apn:details-part-gubser} to be given by\footnote{Formulas for the multiplicity and average transverse momentum of particles produced from Gubser hydrodynamic profiles have also been explored in ref.~\cite{Hatta:2014jva}, for example.}

\begin{subequations}
\label{eq:Nmr-dE/dy-Tfo-expn}
\begin{align}
\label{eq:Nmr-dE/dy-Tfo-expn-1}
N_{\mathrm{mr}} & \approx \frac{4\zeta(3)}{\pi} \Hat{T}_0^3 \left[1  - \frac{1}{2}(q \tau_{0})^{1/2} \left( \frac{T_{\mathrm{FO}}}{q \Hat{T}_{0}}\right)^{3/2} \right],
\\
\label{eq:Nmr-dE/dy-Tfo-expn-2}
N_{\mathrm{mr}} \langle p_{T} \rangle &  \approx
\frac{4 \pi ^{7/2} q \Gamma \left(\frac{11}{6}\right)}{45 \Gamma \left(\frac{7}{3}\right)} \Hat{T}_0^4. 
\end{align}
\end{subequations}
We discuss in Appendix \ref{apn:details-part-gubser} why finite $T_{\mathrm{FO}}$ corrections for 
$N_{\mathrm{mr}}$ are larger than for $N_{\mathrm{mr}} \langle p_{T} \rangle$, hence the results are truncated at different orders in $T_{\mathrm{FO}}/T_0$.

From the above expressions, we derive
\begin{equation}
\label{eq:cs2-observable-2}
\begin{aligned}
\frac{ d\ln \langle p_{T} \rangle}{d \ln N_{\mathrm{mr}}}
&
=
\frac{ d\ln \langle p_{T} \rangle/d \alpha}{d \ln N_{\mathrm{mr}}/d \alpha}
\approx
\frac{1}{3} - \frac{2}{3} \frac{1}{(1 + q^{2} \tau_{0}^{2})} \left( \frac{T_{\mathrm{FO}}}{ T_{0}}\right)^{3/2}
\end{aligned}    
\end{equation}
where the $\alpha$ derivatives are performed by considering the family of rescaled initial temperature profiles $T_{0} = \alpha T_{0}^{\star}$ discussed in Section~\ref{sec:alphafamilona}.
We see that the slope prescription indeed recovers the value of the speed of sound $c_{s}^{2} = 1/3$ for Gubser flow when $T_{\mathrm{FO}} \to 0$. The result does not depend on the normalization constants entering in eq.~\eqref{eq:Nmr-dE/dy-Tfo-expn}.

\begin{figure}[t]
	\centering   
	\begin{subfigure}{0.5\textwidth}
		\includegraphics[scale=0.3]{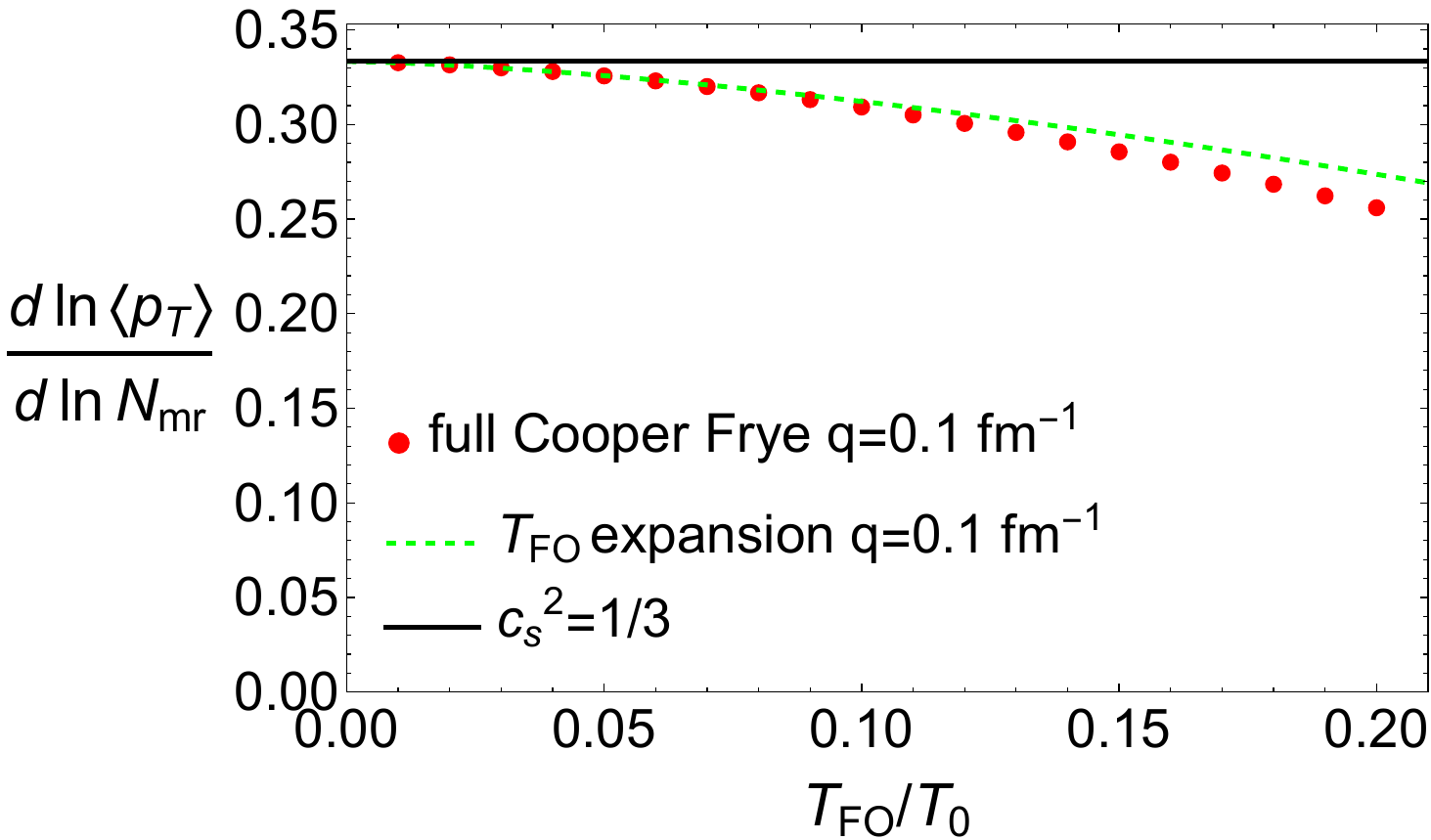}
		\caption{$q=0.1$ fm${}^{-1}$}
		\label{fig:}    
	\end{subfigure}\hfil
	\begin{subfigure}{0.5\textwidth}
		\includegraphics[scale=0.3]{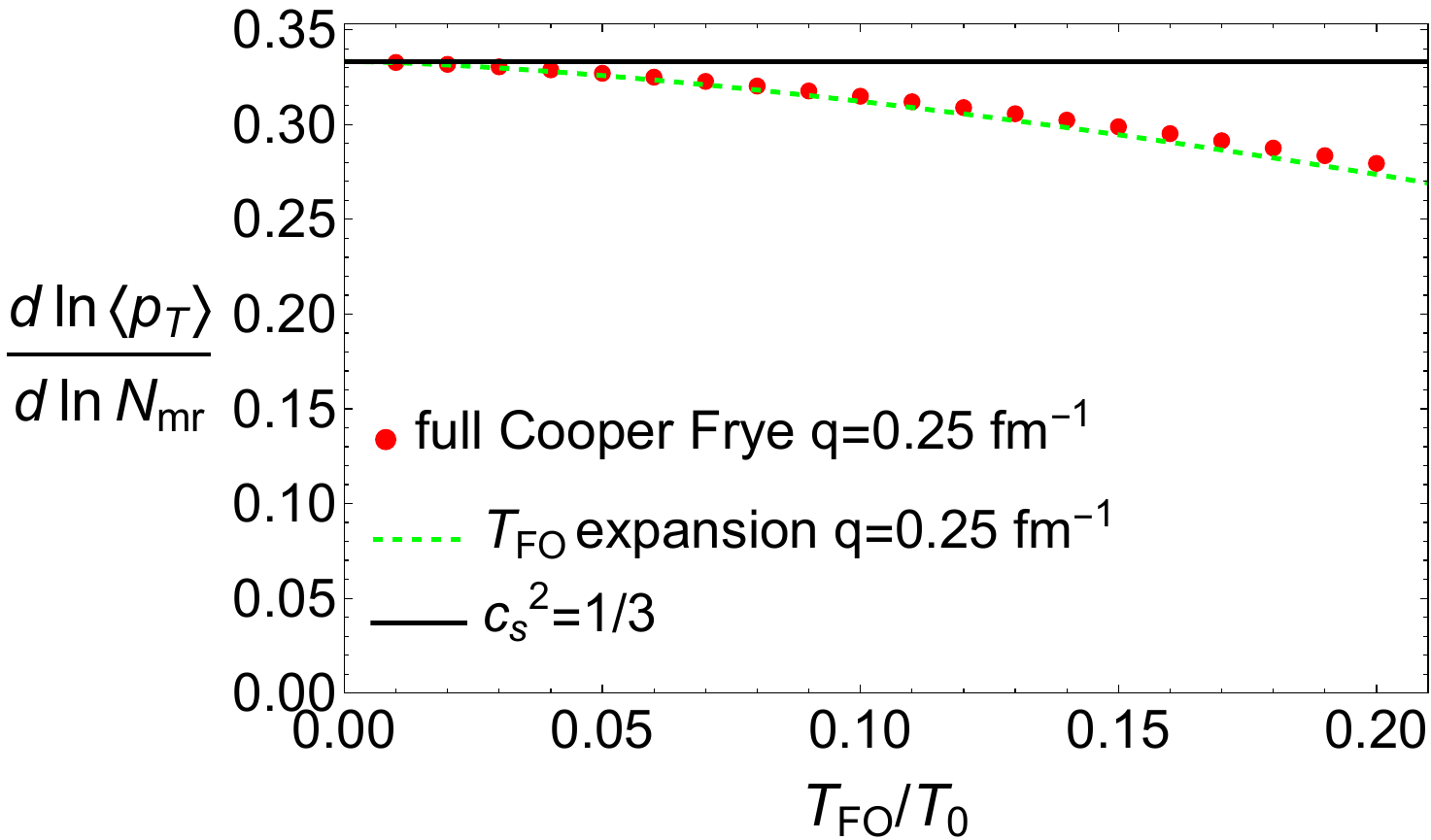}
		\caption{$q=0.25$ fm${}^{-1}$}
		\label{fig:}    
	\end{subfigure}\hfil
        \caption{Comparison between the full particlization formulas (eq.~\eqref{eq:observables-kappa-T} with initial profiles given by eq.~\eqref{eq:gubser-all}) and the low $T_{\mathrm{FO}}$ expansion formula (eq.~\eqref{eq:cs2-observable-2-simpler}). }
	\label{fig:cs2-observable-apprx}
\end{figure}

The leading $T_{\mathrm{FO}}/T_0$ correction depends on the Gubser inverse width  $q$ and initial Milne time $\tau_0$ through the factor $(1+q^2 \tau_0^2)$.
This shows clearly that the midrapidity multiplicity and average transverse momentum do retain a memory of the initial conditions through $q$ and $\tau_{0}$, a memory which fades as $T_{\mathrm{FO}} \to 0$.
This dependence on the initial conditions is a reminder that treating the system as a thermodynamic one is only possible at sufficiently late times; eq.~\eqref{eq:cs2-observable-2} provides a direct quantification of this effect.

Note that in general, the values usually employed for $q$ and $\tau_{0}$ in heavy-ion applications are such that $q \tau_{0}$ is small. 
Equation~\eqref{eq:cs2-observable-2} can thus be approximated well by the simpler expression 
\begin{equation}
\label{eq:cs2-observable-2-simpler}
\begin{aligned}
\frac{ d\ln \langle p_{T} \rangle}{d \ln N_{\mathrm{mr}}}
&
\approx
\frac{1}{3} - \frac{2}{3} \left( \frac{T_{\mathrm{FO}}}{ T_{0}}\right)^{3/2}
.
\end{aligned}    
\end{equation}

In fig.~\ref{fig:cs2-observable-apprx}, we assess the agreement between the full ultra-relativistic Cooper-Frye formulas (eq.~\eqref{eq:observables-kappa-T} with profiles given by eq.~\eqref{eq:gubser-all}) and expression \eqref{eq:cs2-observable-2-simpler} in the range $ 0 \leq T_{\mathrm{FO}}/T_0 \leq 0.2$ GeV for two different values of the Gubser inverse width $q$. We see that eq.~\eqref{eq:cs2-observable-2-simpler} captures very well the $T_{\mathrm{FO}}/T_0$ of the slope observable, with the numerical calculations (points) showing only slight deviations from the analytical formula (dashed lines).

\section{Summary}
\label{eq:conclusion}

In this work, we studied the relation between the speed of sound of a fluid and its final particle multiplicity and average energy.
Focusing on inviscid hydrodynamics with a constant speed of sound, we studied initial conditions where the temperature is scaled by an overall constant without change in its shape, a simplified model of ultracentral heavy-ion collisions.

Within this model, the average particle energy depends on the particle multiplicity strictly as a power law, with the power set by the square of the speed of sound, 
$E_{\textrm{tot}}/N_{\textrm{tot}}\propto N_{\textrm{tot}}^{c_s^2}$, but only if the total energy and multiplicity of all particles can be measured.
When restricting the set of measured final particles, the relation between the multiplicity, the energy and the speed of sound is only recovered for an infinitely long-lived fluid, which we characterized by an infinitely low freezeout temperature $T_{\mathrm{FO}} \to 0$. Using three-dimensional hydrodynamic simulations,  we confirmed the effect of rapidity cuts or transverse momentum cuts on the relation between the average energy $E/N$ and the multiplicity $N$. We found a non-trivial dependence on rapidity due to the early freezeout of large-rapidity fluid elements when a fixed-temperature freezeout criteria is used.

To study in greater detail the midrapidity case, we used a simpler model with longitudinal boost invariance and cylindrical symmetry in the transverse plane. We focused on massless bosons produced at midrapidity, where the average energy corresponds to the average transverse momentum.
We saw in this model that $d\ln \langle p_{T} \rangle/d \ln N_{\mathrm{mr}}$ converges  as expected  to $c_{s}^{2}$ for asymptotically low freezeout temperatures. For a fixed ratio of the freezeout temperature to plasma's initial maximum temperature, $T_{\mathrm{FO}}/T_0$, we saw that the convergence of $d\ln \langle p_{T} \rangle/d \ln N_{\mathrm{mr}}$ to $c_{s}^{2}$ is faster for smaller $c_s^2$. 

Building on the above model, we used Gubser hydrodynamic solution to obtain expressions for $d\ln \langle p_{T} \rangle/d \ln N_{\mathrm{mr}}$. The resulting formula quantifies explicitly the dependence of the latter quantity on the initial conditions and the freezeout temperature of the system, clarifying how and when the system approaches a ``thermodynamic'' limit where only conserved quantities are relevant.

While our model of ultra-relativistic collisions is a simplified one, our results suggest that the speed of sound will be best accessed with the highest possible center-of-mass energy collisions, where $T_{\mathrm{FO}}/T_0$ corrections will be smallest. Conversely, using eq.~\eqref{eq:cs2-observable} in lower energy collisions may lead to inaccurate extractions of the speed of sound if the plasma's initial temperature is too low.
Moreover, the non-trivial dependence on rapidity cuts observed in our model indicates future studies should be vigilant for this effect.

An important next step is to study temperature-dependent speeds of sound, which is the relevant scenario for nuclear matter. The effective temperature entering in eq.~\eqref{eq:cs2-observable} is an essential ingredient of the original formulation of the slope observable in ref.~\cite{Gardim:2019xjs}, and the constant speed of sound model employed in the present work does not provide insights into the role of this parameter. In a different direction, it will be relevant to extend our reasoning to dissipative fluids and more general initial conditions. These extensions will depend more on numerical insights than analytical ones, and will be the subject of separate publications.

\section*{Acknowledgments}

The authors thank Sangyong Jeon, Andi Mankolli, Shengquan Tuo, Julia Velkovska and an anonymous referee for useful discussions. G.~S.~R.~, M.~S.~and J.-F.~P.~are supported by Vanderbilt University and by the U.S. Department of Energy, Office of Science under Award Number DE-SC-0024347. L.G. is partially supported by a Vanderbilt Seeding Success grant. This research used resources of the National Energy Research Scientific Computing Center, a DOE Office of Science User Facility supported by the Office of Science of the U.S. Department of Energy
under Contract No. DE-AC02-05CH11231 using NERSC award NP-ERCAP0029957.

\appendix

\section{Details for the particlization formulas}
\label{apn:details-part}

In this Appendix\footnote{A repository containing the notebooks with the derivations in this appendix as well as related figures is available at \url{https://github.com/gabriel-sr151/cylindrical-sym-CF} .}, we shall provide further details on the derivation of eqs.~\eqref{eq:observables-kappa-T-Nmr} and \eqref{eq:observables-kappa-T}. In what follows, we parametrize the particle momentum $p^{\mu}$ so that the transverse momentum 3-vector, $\Vec{p}_{T}$, points in the $x$-direction, thus $(p^{\mu})_{t,x,y_{s},z} = (m_{T} \cosh y,  p_{T} , 0 , m_{T} \sinh y)$, where $m_{T} = \sqrt{p^{2}_{T} + m^{2}}$, which reduces to $m_{T} = p_{T}$ for massless particles, $y$ is the particle momentum-rapidity. In this case, the hypersurface element in cylindrical coordinates reads $(d \Sigma^{\mu})_{t,r,\phi,z} = ( r dr d \phi dz, r d\phi dz dt, 0 , r d \phi dr dt)$. For a cylindrically-symmetric, longitudinally boost-invariant flow, the fluid 4-vector can be parametrized as $(u^{\mu})_{t,r,\phi,z} = \left( \cosh \kappa \cosh \eta_{s} , \sinh \kappa \cosh \eta_{s}, 0, \sinh \eta_{s} \right)$. Then, we have that $u_{\mu} p^{\mu} = m_{T} \cosh \kappa \cosh(\eta_{s} -y) - p_{T} \sinh \kappa \cos \phi$. Besides,  
\begin{align*}
d \Sigma^{\mu} p_{\mu} = d\phi d\eta_{s} r [m_{T} \tau \cosh (\eta_{s}-y) dr + m_{T} \sinh (\eta_{s}-y) d \tau - p_{T} \cos \phi d \tau],
\end{align*}
but, since we shall integrate $\eta_{s} \in (- \infty, +\infty)$, the second term of this expression does not contribute. Then, we have for the ultrarelativistic ($m_{T} \approx p_{T}$) particle yield at midrapidity ($y=0$),
\begin{equation}
\begin{aligned}
N_{\mathrm{mr}}
& =
\left. 
\frac{dN}{dy} \right\vert_{y=0}
=
\left. 
\int d^{2}p_{T} \frac{dN}{dy d^{2}p_{T}}
\right\vert_{y=0}
\\
&
=
 \frac{2 \pi  g}{(2 \pi )^{3}}
 \int_{\Gamma} \int_{0}^{2 \pi} d\phi \int_{\mathds{R}} d \eta_{s} \int_{0}^{\infty} dp_{T} p_{T}^{2}   \frac{\tau  r [\cosh \eta_{s} dr - \cos \phi d \tau]}
 {\exp [p_{T} ( \cosh \kappa \cosh \eta_{s} - \sinh \kappa \cos \phi)/T_{\mathrm{FO}}] - 1}
\\
&
=
 \frac{g T^{3}_{\mathrm{FO}}\zeta(3)}{2 \pi^{2}}
 \int_{\Gamma} \int_{\mathds{R}} d \eta_{s}  \int_{0}^{2 \pi} d\phi  \frac{\tau  r [\cosh \eta_{s} dr - \cos \phi d \tau]}
 {(\cosh \kappa \cosh \eta_{s} -\sinh \kappa \cos \phi)^{3}}
\\
&
=
 \frac{g T^{3}_{\mathrm{FO}}\zeta(3)}{\pi^{2}}
 \int_{\Gamma} \int_{\mathds{R}} d \eta_{s}  \int_{0}^{1} \frac{du}{\sqrt{1-u^{2}}}  \frac{\tau  r [\cosh \eta_{s} dr - u d \tau]}
 {(\cosh \kappa \cosh \eta_{s} - u \sinh \kappa )^{3}} 
\\
&
=
 \frac{g T^{3}_{\mathrm{FO}}\zeta(3)}{\pi}
 \int_{\Gamma} \int_{\mathds{R}} d \eta_{s}  \frac{\tau  r \sech^{3} \kappa  [\left( \tanh ^2\kappa  \text{ sech}^4\eta_s+2 \text{ sech}^2\eta _s \right) dr - 3 \tanh \kappa  \text{ sech}^4\eta _s d \tau]}
 {2 \left(1-\sech^2\eta_{s} \tanh^2\kappa \right)^{5/2}}
 \\
&
=
 \frac{2 g T^{3}_{\mathrm{FO}}\zeta(3)}{\pi}
 \int_{\Gamma} \int_{0}^{1} dv
\frac{\tau r [dr \left(v \tanh ^2\kappa +2\right)-d\tau  3 v \tanh \kappa ]}{2 v \sqrt{1-v} \left(1-v \tanh ^2\kappa\right)^{5/2}} 
\\
&
=
 \frac{2g T^{3}_{\mathrm{FO}} \zeta(3)}{\pi}
 \int_{\Gamma}
 \tau r (dr \cosh \kappa  - d\tau \sinh \kappa) 
\end{aligned}    
\end{equation}
where, from the first to the second line we have already performed the angular integral in the transverse momentum 3-vector, $\Vec{p}_{T}$, from the second to third line, we used the fact that $\int_{0}^{\infty}  dx \ x^{2} (e^{A x} - 1)^{-1} = [2 \zeta(3)]/A^{3}$. From the third to the fourth line, we employ the variable transformation $u = \cos \phi$ and perform the corresponding integrals. From the fourth to the fifth line, we perform the variable transformation $v = \sech^{2}\eta_{s}$, which is performed in the step from the fifth to the sixth line. With analogous steps, we can also compute the mean transverse momentum as 
\begin{equation}
\begin{aligned}
N_{\mathrm{mr}} \langle p_{T} \rangle
& 
=
\left. 
\int d^{2}p_{T} \ p_{T} \frac{dN}{dy d^{2}p_{T}}
\right\vert_{y=0}
\\
&
=
 \frac{2 \pi g}{(2 \pi )^{3}}
 \int_{\Gamma} \int_{0}^{2 \pi} d\phi \int_{\mathds{R}} d \eta_{s} \int_{0}^{\infty} dp_{T} p_{T}^{3}   \frac{\tau  r [\cosh \eta_{s} dr - \cos \phi d \tau]}
 {\exp [(m_{T} \cosh \kappa \cosh \eta_{s} - p_{T} \sinh \kappa \cos \phi)/T_{\mathrm{FO}}] - 1}
\\
&
=
 \frac{g T^{4}_{\mathrm{FO}}}{(2 \pi)^{3}}
\frac{2 \pi^{5}}{15}
 \int_{\Gamma} \int_{0}^{2 \pi} d\phi \int_{\mathds{R}} d \eta_{s}  \frac{\tau  r [\cosh \eta_{s} dr - \cos \phi d \tau]}
 {(\cosh \kappa \cosh \eta_{s} -\sinh \kappa \cos \phi)^{4}} 
\\
&
=
 \frac{g T^{4}_{\mathrm{FO}}}{(2 \pi)^{3}}
\frac{2\pi^{6}}{15}
 \int_{\Gamma} \int_{\mathds{R}} d \eta_{s}  \frac{\tau  r \sech^4\kappa \sech^3\eta_{s} }
 {\left(1-\sech^2\eta_{s} \tanh^2\kappa \right)^{7/2}} 
\left[ \left(3 \sech^2\eta_{s}  \tanh^2\kappa + 2 \right) dr 
\right.
\\
&
\left.
-  \left(\sech^4\eta  \tanh^3\kappa +4 \sech^2\eta_{s}  \tanh \kappa \right) d \tau
\right]
\\
&
=
 \frac{g T^{4}_{\mathrm{FO}}}{(2 \pi)^{3}}
\frac{2\pi^{6}}{15}
 \int_{\Gamma}\int_{0}^{1} du \frac{\tau  r \sech^4\kappa \ u^{3/2} [ \left(3 u  \tanh^2\kappa + 2 \right) dr -  \left(u^{2}  \tanh^3\kappa +4 u  \tanh \kappa \right) d \tau]}
 {2 u \sqrt{1-u} \left(1-u \tanh^2\kappa \right)^{7/2}} 
 \\
&
=
\frac{g T^{4}_{\mathrm{FO}}}{( 2 \pi )^{3}}
\frac{4\pi^{6}}{45}
\int_{\Gamma} \ \tau r \left\{\left[ (4 \cosh (2 \kappa)+3) \text{sech}^2\kappa E\left(-\sinh ^2\kappa\right)-4 K\left(-\sinh ^2\kappa\right) \right] \cosh \kappa \ dr 
\right.
\\
&
\left.
-
\csch^2\kappa  \left[ 
(1-2 \cosh (2 \kappa)) K\left(-\sinh ^2\kappa\right)+(4 \cosh (2 \kappa)-3) E\left(-\sinh ^2\kappa\right)
\right]\sinh \kappa \ d\tau \right\} ,
\end{aligned}    
\end{equation}
where, from the second to third line, we used the fact that $\int_{0}^{\infty} dx \ x^{3} (e^{A x} - 1)^{-1} = [\pi^{4}/(15A^{4})]$.

\section{Gubser flow particlization and the low $T_{\mathrm{FO}}$ limit}
\label{apn:details-part-gubser}

\begin{figure}[!h]
\centering   
\begin{subfigure}{0.5\textwidth}
    \includegraphics[scale=0.28]{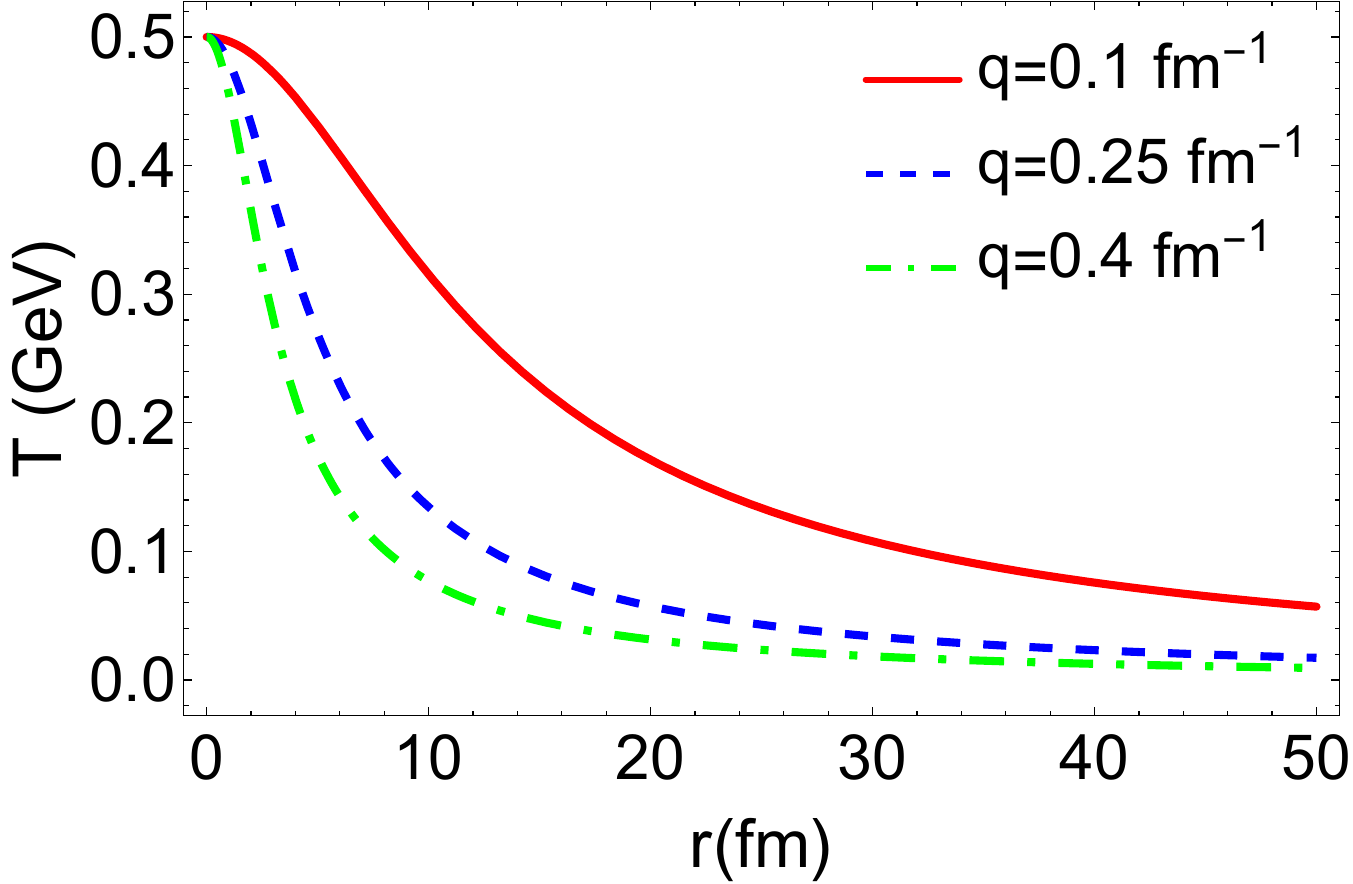}
    \caption{Initial temperature profiles at $\tau = \tau_{0}$}
    \label{fig:Gubser_init_T}    
\end{subfigure}\hfil
\begin{subfigure}{0.5\textwidth}
    \includegraphics[scale=0.28]{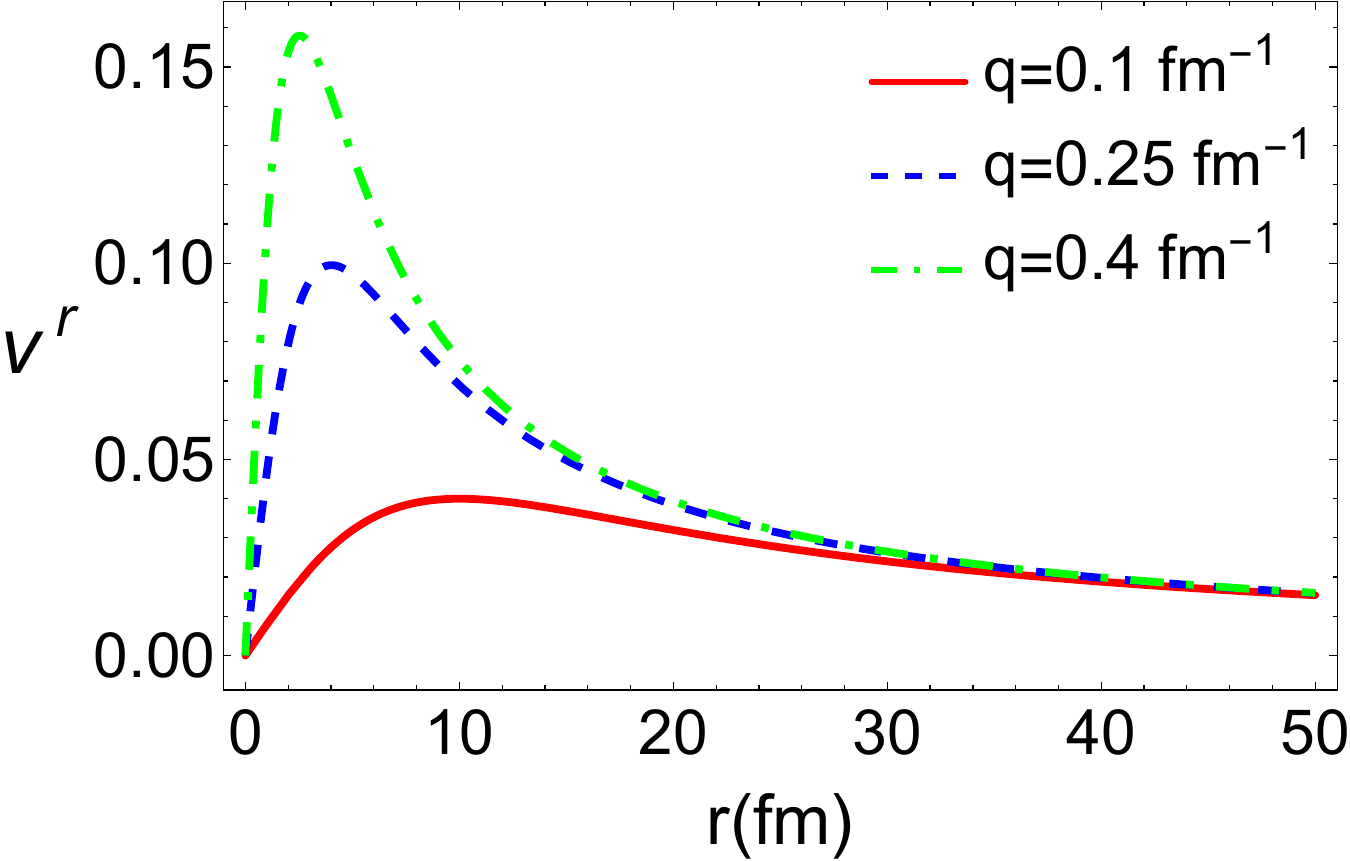}
    \caption{Initial velocity profiles at $\tau = \tau_{0}$}
    \label{fig:Gubser_init_vr}    
\end{subfigure}\hfil
\begin{subfigure}{0.5\textwidth}
    \includegraphics[scale=0.29]{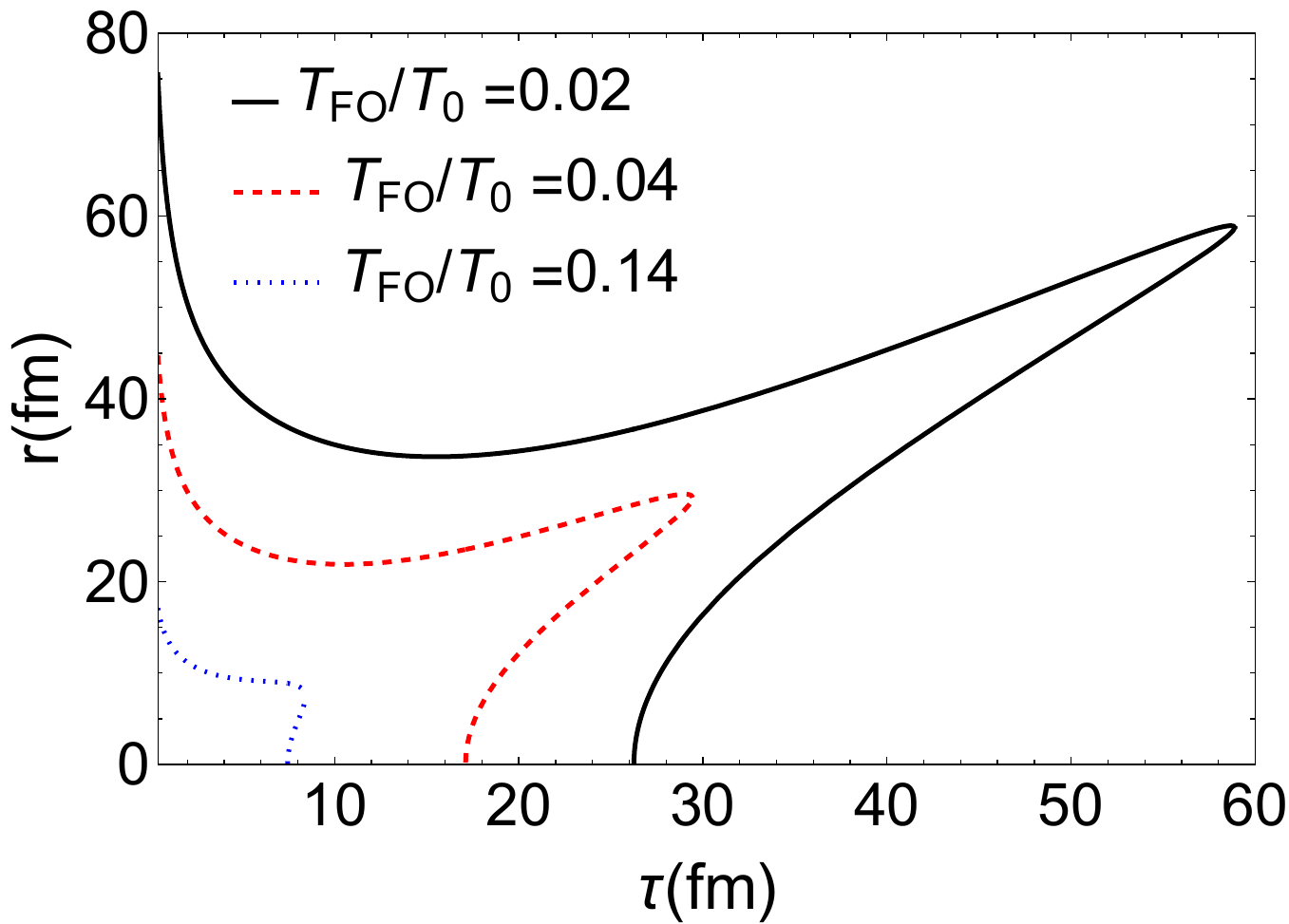}
    \caption{Low $T_{\mathrm{FO}}$ profiles $q=0.25$ fm${}^{-1}$.}
    \label{fig:dln<pT>/dlnN-gubser-4}    
\end{subfigure}\hfil
\caption{Gubser flow initial profile, freezeout temperature profiles and speed-of-sound estimates for $\tau_{0} = 0.4$ fm $T_{0} = 0.5$ GeV. }
\label{fig:T-obsv-gubser}
\end{figure}

For small values of the parameter $q$, the Gubser solution \eqref{eq:gubser-all} for inviscid fluids exhibits initial temperature and flow profiles that share many similarities to the initial conditions expected in ultrarelativistic heavy-ion collisions: monotonically decreasing temperature profile in the radial direction with a peak at $r=0$, and larger flow velocities in regions of larger initial temperature gradients. These features are illustrated in figs.~\ref{fig:Gubser_init_T} and \ref{fig:Gubser_init_vr} for three different small values of $q$, at an initial time $\tau = \tau_{0} = 0.4$ fm.

In fig.~\ref{fig:dln<pT>/dlnN-gubser-4}, we see, similarly with Section~\ref{sec:cilindro}, that the low $T_{\mathrm{FO}}$ hypersurfaces are sharply peaked. In the present case, this occurs exactly at $\tau = r$. 

\begin{figure}[!h]
\centering
    \includegraphics[scale=0.3]{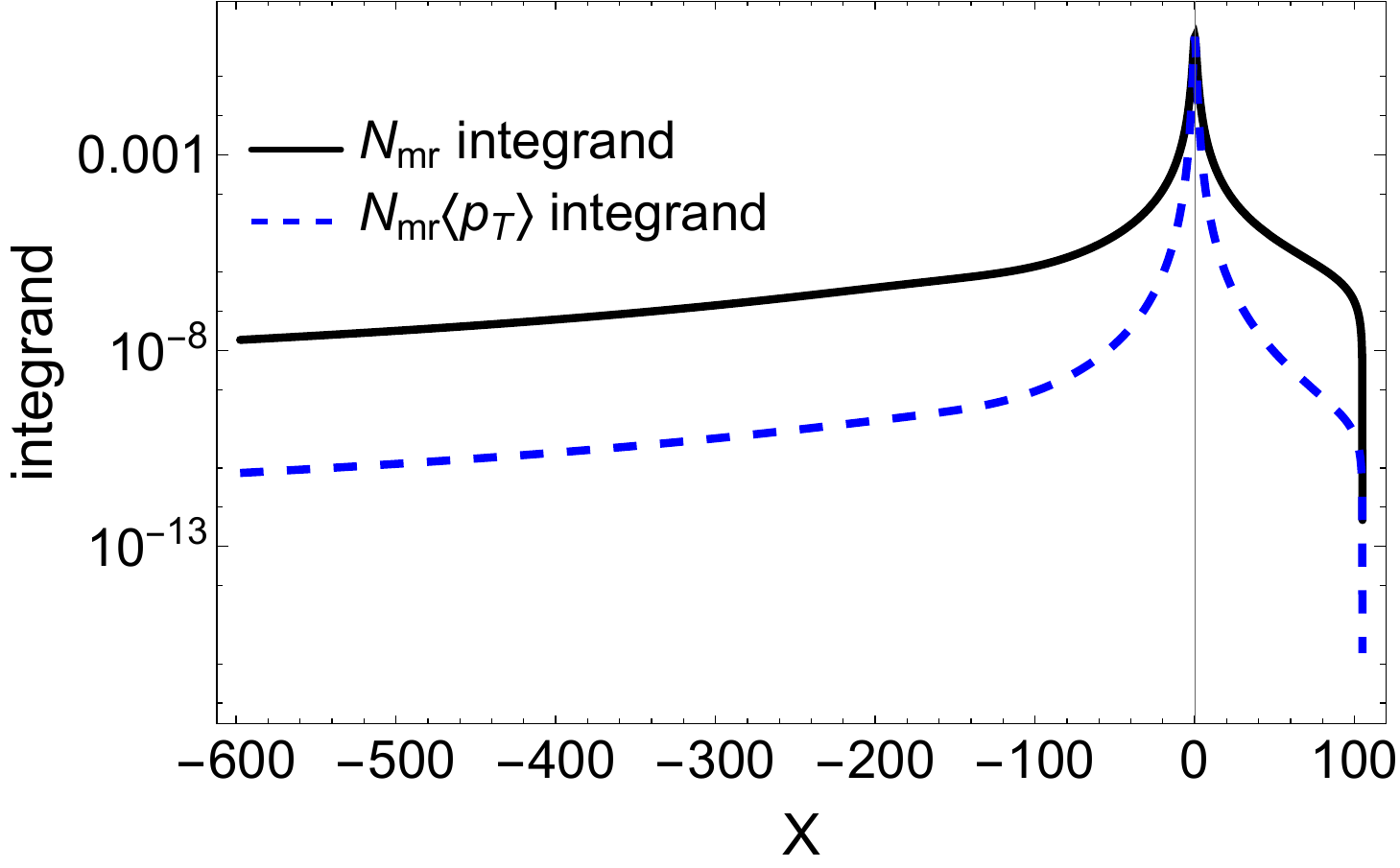}
\caption{Multiplicity and mean transverse momentum integrands (normalized with their corresponding values in $X = q(\tau - r)= 0$) for $T_{\mathrm{FO}}/T_{0} = 2 . 10^{-4}$ plotted in the interval $(X_{-}, X_{+})$ where the integration takes place.}
\label{fig:integrands}
\end{figure}

In the  $T_{\mathrm{FO}} \to 0$ limit, the multiplicity $N_{\mathrm{mr}}$ and transverse momentum $N_{\mathrm{mr}} \langle p_{T} \rangle$ are dominated by contributions originating from near the $r = \tau$ line. This can be seen when plotting the integrands of eqs.~\eqref{eq:observables-kappa-T-Nmr} and \eqref{eq:observables-kappa-<pt>} in terms of the dimensionless coordinates $X = q(\tau - r)$, $Y = q(\tau + r)$, instead of $\tau$ and $r$, such that $X$ near $0$ corresponds to $\tau \approx r$. In fig.~\ref{fig:integrands}, we plot
$\cosh \kappa \ (dr/dX) 
-
\sinh \kappa \ (d\tau/dX)$ (the $N_{\mathrm{mr}}$ integrand)
and 
 $F(\kappa) \cosh \kappa\ (dr/dX) 
-
 G(\kappa) \sinh \kappa \ (d\tau/dX)$ (the $N_{\mathrm{mr}} \langle p_{T} \rangle$ integrand) as a function of $X$, where $\kappa = \kappa(X, Y(X))$ is the fluid rapidity at the freezeout surface defined by $Y(X)$, which is defined by implicitly the freezeout contour 
 \begin{align*}
T\left(\tau=\frac{X+Y}{2 q},r=\frac{Y-X}{2 q}\right) = T_{\mathrm{FO}}
 \end{align*}
Figure~\ref{fig:integrands} also shows that the $N_{\mathrm{mr}} \langle p_{T} \rangle$ integrand is much more sharply peaked around $X = 0$ than the $N_{\mathrm{mr}}$ integrand. This will have an effect on how we perform the $T_{\mathrm{FO}} \to 0$ expansion. 

In $(X,Y)$ coordinates, the multiplicity and mean transverse momentum integrals become 
\begin{equation}
\begin{aligned}
&
N_{\mathrm{mr}} 
=
 \frac{2g T^{3}_{\mathrm{FO}}\zeta(3)}{\pi}
\left(\frac{\Hat{T}_{0}}{2 q T_{\mathrm{FO}}} \right)^{3/2} \int_{X_{-}}^{X_{+}}
 dX \ \frac{(Y-X) (Y+X)^{1/2}(3 X^2+4 X Y+3 Y^2+2)}{\left(X^2+1\right) \left(2 X Y+3 Y^2+1\right)},
\\
&
N_{\mathrm{mr}} \langle p_{T} \rangle
=
\frac{g T^{4}_{\mathrm{FO}}}{( 2 \pi )^{3}}
\frac{4\pi^{6}}{45}  
\frac{1}{16 q^{3}}
\left(\frac{\Hat{T}_{0}}{2 q T_{\mathrm{FO}}} \right)^{3/2}
\int_{X_{-}}^{X_{+}}
 dX \ \frac{24 (Y-X) (X+Y)^{3/2} (1+Y^{2})}{\left(X^2+1\right) \left(X^2+Y^2+2\right) \left(2 X Y+3 Y^2+1\right)}
\\
&
\times
\left[
\left(
2 X^4+2 X^3 Y+X^2 \left(3 Y^2+5\right)+2 X Y \left(Y^2+2\right)+2 Y^4+5 Y^2+3
\right)
 E\left(-\frac{\left(X^2-Y^2\right)^2}{4 \left(X^2+1\right) \left(Y^2+1\right)}\right)
\right. 
\\
&
\left. 
-
\left( X^4+X^3 Y+X^2 \left(2 Y^2+3\right)+X Y \left(Y^2+2\right)+Y^4+3 Y^2+2 \right)
K\left(-\frac{\left(X^2-Y^2\right)^2}{4 \left(X^2+1\right) \left(Y^2+1\right)}\right)
\right]
\end{aligned}    
\end{equation}
where $Y = Y(X)$ given implicitly by $Y^2 \left(X^3 T_{\text{FO}}^3+X T_{\text{FO}}^3\right)+X^3 T_{\text{FO}}^3+Y^3 \left(X^2 T_{\text{FO}}^3+T_{\text{FO}}^3\right)+Y \left(X^2 T_{\text{FO}}^3+T_{\text{FO}}^3\right)+X T_{\text{FO}}^3-8 q^3 \hat{T}_0^3 = 0$. The integration limits $X_\pm$ are obtained from the original $\tau,r$ contour limits as follows: $X_+$ corresponds to the boundary at $r=0$ and $X_-$ corresponds to the boundary at $\tau=\tau_{0}$. More explicitly, $X_+$  is the real (positive) solution of the polynomial equation $-4 q^3 \Hat{T}_{0}^3+T_{\mathrm{FO}}^3 X_+^5+2 T_{\mathrm{FO}}^3 X_+^3+T_{\mathrm{FO}}^3 X_+ = 0$ and $X_-$  is the real (negative) solution of the polynomial equation $\tau_{0} T_{\text{FO}}^3 -4 q^2 \Hat{T}_{0}^3+X_-^2 \left(4 q^2 \tau_{0}^3 T_{\text{FO}}^3+2 \tau_{0} T_{\text{FO}}^3\right)+4 q^2 \tau_{0}^3 T_{\text{FO}}^3-4 q \tau_{0}^2 T_{\text{FO}}^3 X_-^3-4 q \tau_{0}^2 T_{\text{FO}}^3 X_-+\tau_{0} T_{\text{FO}}^3 X_-^4 = 0$. 

In the limit where $T_{\text{FO}} \to 0$, we have that 
\begin{equation}
\begin{aligned}
X_- & \approx - \sqrt{2}[(\hat{T}_{0}^{3}q^{2})/(\tau_{0} T_{\mathrm{FO}}^{3})]^{1/4} \to -\infty,
\quad
X_+ \approx 2^{2/5} [(\hat{T}_{0}^{3}q^{3})/(T_{\mathrm{FO}}^{3})]^{1/5} \to +\infty,
\\
Y(X) & \approx \frac{2 q \Hat{T}_{0}}{T_{\mathrm{FO}}} \frac{1}{(1+X^{2})^{1/3}} \to + \infty ,
\end{aligned}
\end{equation}
where the arrows denote the limit of the corresponding quantity in the strict limit as $T_{\mathrm{FO}} \to 0$. Thus, this limit amounts to taking the integration boundaries to $\pm \infty$ and expanding the integrand in an asymptotic series as $Y(X) \to \infty$. At leading order, the multiplicity integral becomes
\begin{equation}
\label{eq:LO-nmr}
\begin{aligned}
N_{\mathrm{mr}} 
&=
 \frac{2g T^{3}_{\mathrm{FO}}\zeta(3)}{\pi}
 \left(\frac{\Hat{T}_{0}}{2 q T_{\mathrm{FO}}} \right)^{3/2}
 \int_{X_-}^{X_+}
 dX \ \frac{Y(X)^{3/2}}{(1+X^{2})} + \mathcal{O}\left( T_{\mathrm{FO}} \right)
\\
&
=
 \frac{2g T^{3}_{\mathrm{FO}}\zeta(3)}{\pi}  \int_{-\infty}^{+\infty}
 dX \ \frac{1}{(1+X^{2})^{3/2}} + \mathcal{O}\left( T_{\mathrm{FO}} \right) \\
 & 
 =
  \frac{4 g \Hat{T}^{3}_{0}\zeta(3)}{\pi} + \mathcal{O}\left( T_{\mathrm{FO}} \right).
\end{aligned}    
\end{equation}

Similarly, for the transverse momentum, we find
\begin{equation}
\label{eq:LO-energy}
\begin{aligned}
N_{\mathrm{mr}} \langle p_{T} \rangle
&=
\frac{g T^{4}_{\mathrm{FO}}}{( 2 \pi )^{3}}
\frac{4\pi^{6}}{45}  
\frac{1}{16 q^{3}}
\left(\frac{\Hat{T}_{0}}{2 q T_{\mathrm{FO}}} \right)^{3/2}
\int_{X_{-}}^{X_{+}}
 dX \frac{8 Y(X)^{11/2}}{\sqrt{X^2+1}} + \mathcal{O}\left( T_{\mathrm{FO}} \right)
\\
&
=
\frac{g}{( 2 \pi )^{3}}
\frac{4\pi^{6}}{45}  
q
\Hat{T}_{0}^{4}
\int_{-\infty}^{\infty}
 dX \frac{8}{\left(X^2+1\right)^{7/3}} + \mathcal{O}\left( T_{\mathrm{FO}} \right)
\\
&
=
\frac{4 \pi ^{7/2} q g \Gamma \left(\frac{11}{6}\right)}{45 \Gamma \left(\frac{7}{3}\right)} \Hat{T}_0^4 + \mathcal{O}\left( T_{\mathrm{FO}} \right).
\end{aligned}  
\end{equation}

Calculating $\mathcal{O}\left( T_{\mathrm{FO}} \right)$ corrections to the multiplicity and transverse momentum is considerably more challenging.
A key simplification in the eqs.~\eqref{eq:LO-nmr} and ~\eqref{eq:LO-energy} is the fact that the integrand is dominated by the $X=0$ regime where $Y$ is very large. Corrections of order $\mathcal{O}\left( T_{\mathrm{FO}} \right)$ originate from regions where $X$ is not small and $Y$ is not large, making series expansion more challenging.
On the other hand, as argued above with fig.~\ref{fig:integrands}, since the integrand of $N_{\mathrm{mr}}\langle p_{T} \rangle$ is more sharply peaked around $X = 0$, we do expect smaller $\mathcal{O}\left( T_{\mathrm{FO}} \right)$ corrections; we verified this conclusion numerically. It is thus sufficient to find the $\mathcal{O}\left( T_{\mathrm{FO}} \right)$ to the multiplicity, eq.~\eqref{eq:LO-nmr}.

\begin{figure}[!h]
\centering   
\begin{subfigure}{0.5\textwidth}
    \includegraphics[scale=0.29]{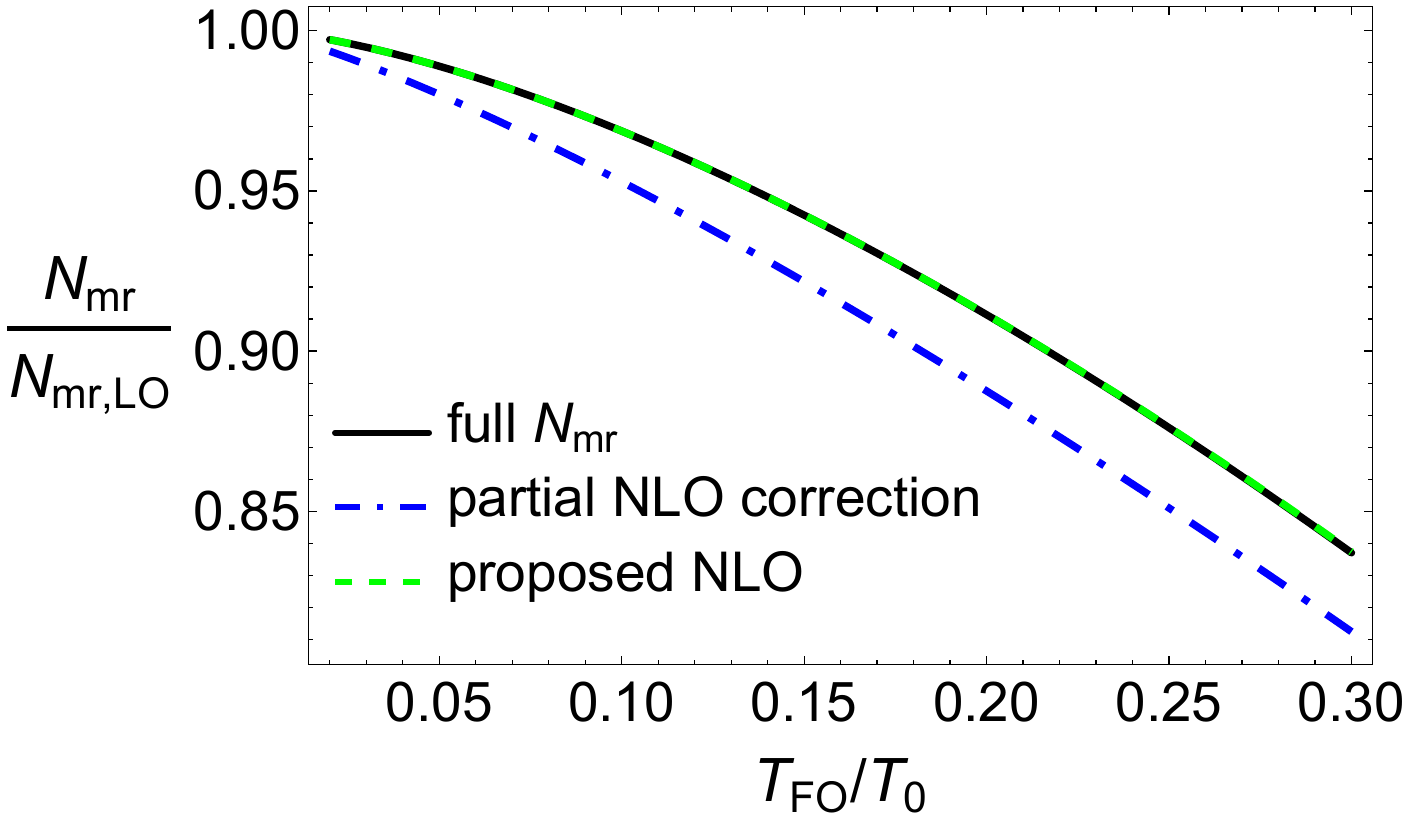}
    \caption{Comparison between NLO expressions, $q=0.1$.}
    \label{fig:gubser-nmr-compar-NLO}    
\end{subfigure}\hfil
\begin{subfigure}{0.5\textwidth}
    \includegraphics[scale=0.30]{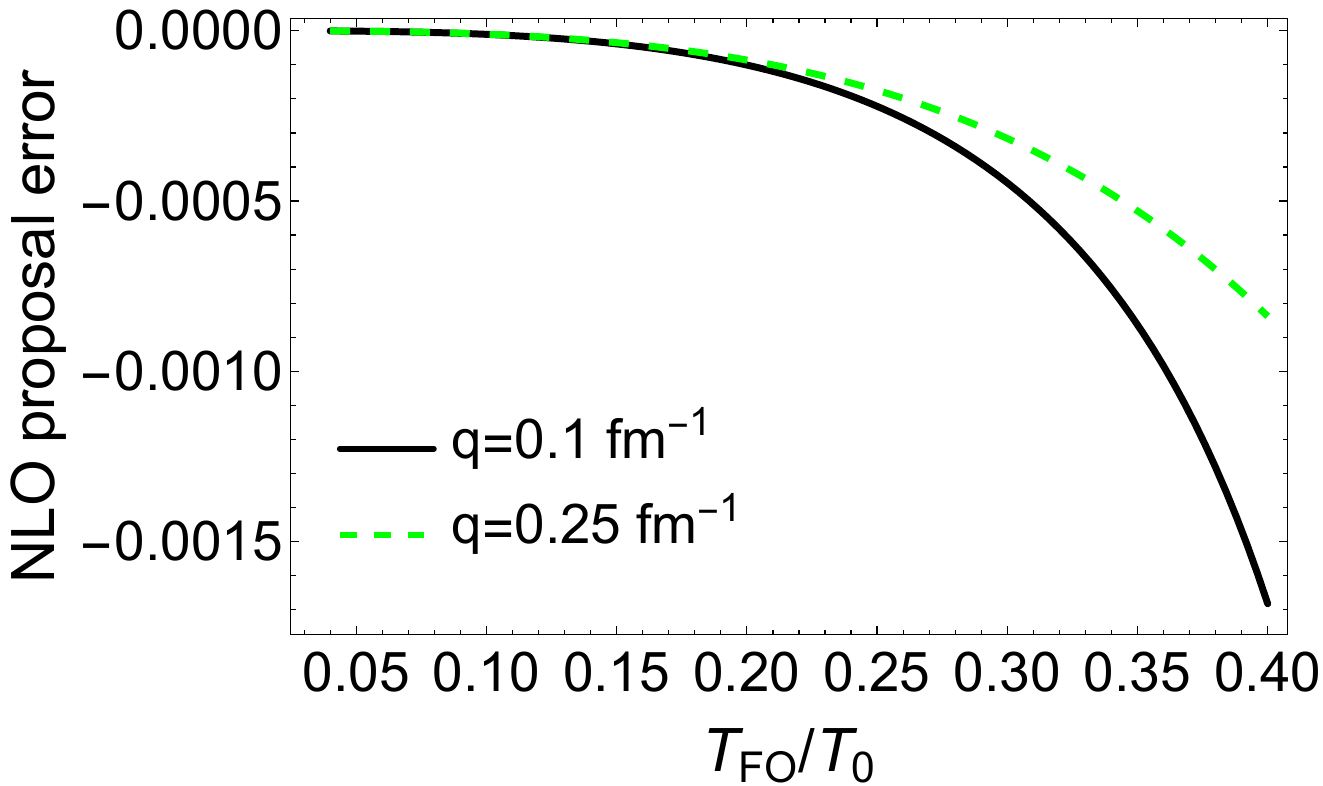}
    \caption{Relative error of expression \eqref{eq:nmr-NLO-prop} compared to the exact numerical result.}
    \label{fig:gubser-nmr-NLO-err}    
\end{subfigure}\hfil

\caption{Numerical tests for the proposed NLO $T_{\mathrm{FO}} \to 0$ expression correction for $N_{\mathrm{mr}}$. (a) Comparison between eq.~\eqref{eq:nmr-NLO-prop}, the ``proposed NLO'' and eq.~\eqref{eq:nmr-NLO-syst}, the ``partial NLO correction''. (b) Relative error of the NLO expression proposal \eqref{eq:nmr-NLO-prop}.}
\label{fig:nmr-prop}
\end{figure}

Guidance on the $\mathcal{O}\left( T_{\mathrm{FO}} \right)$ to the multiplicity can be obtained from investigating corrections stemming from the fact that the integration interval is finite. One obtains
\begin{equation}
\label{eq:nmr-NLO-syst}
\begin{aligned}
N_{\mathrm{mr}} 
&\approx
 \frac{2g T^{3}_{\mathrm{FO}}\zeta(3)}{\pi}
 \left(\frac{\Hat{T}_{0}}{2 q T_{\mathrm{FO}}} \right)^{3/2}
 \int_{X_-}^{X_+}
 dX \ \frac{Y(X)^{3/2}}{(1+X^{2})}
\\
&
\approx
 \frac{2g T^{3}_{\mathrm{FO}}\zeta(3)}{\pi}  \int_{X_-}^{X_+}
 dX \frac{1}{(1+X^{2})^{3/2}}  
\\
&
\approx
  \frac{4 g \Hat{T}^{3}_{0}\zeta(3)}{\pi}
  \left[ 1 - \frac{1}{2^{14/5}} \left(\frac{T_{\mathrm{FO}}}{q \Hat{T}_{0}} \right)^{6/5} - \frac{1}{8}(q \tau_{0})^{1/2} \left( \frac{T_{\mathrm{FO}}}{q \Hat{T}_{0}}\right)^{3/2} \right].
\end{aligned}    
\end{equation}
The above expression is not a systematic expansion\footnote{Effects of order $\mathcal{O}\left( T_{\mathrm{FO}} \right)$ also stem from corrections to the integrand.}  in $T_{\mathrm{FO}}$, but it does capture some of the dependence of the multiplicity on $T_{\mathrm{FO}}$.
Based on this insight and on numerical tests, 
we found that the following expression captures the $T_{\mathrm{FO}}$ dependence of $N_{\mathrm{mr}}$ to very high accuracy:
\begin{equation}
\label{eq:nmr-NLO-prop}
\begin{aligned}
N_{\mathrm{mr}} & \approx \frac{4g\zeta(3)}{\pi} \Hat{T}_0^3 \left[1  - \frac{1}{2}(q \tau_{0})^{1/2} \left( \frac{T_{\mathrm{FO}}}{q \Hat{T}_{0}}\right)^{3/2} \right].
\end{aligned}    
\end{equation}
Agreement with numerical results is shown in fig.~\ref{fig:nmr-prop}, where eq.~\eqref{eq:nmr-NLO-prop} is labeled as ``proposed NLO'' and eq.~\eqref{eq:nmr-NLO-syst} is labeled as ``partial NLO correction'', both normalized with the LO expression \eqref{eq:LO-nmr}. In fact, in panel \ref{fig:gubser-nmr-NLO-err}, we see that the error of proposal \eqref{eq:nmr-NLO-prop} is at most $0.2\%$ in the interval considered.

\bibliography{liography}

\begin{thebibliography}{48}%
\makeatletter
\providecommand \@ifxundefined [1]{%
 \@ifx{#1\undefined}
}%
\providecommand \@ifnum [1]{%
 \ifnum #1\expandafter \@firstoftwo
 \else \expandafter \@secondoftwo
 \fi
}%
\providecommand \@ifx [1]{%
 \ifx #1\expandafter \@firstoftwo
 \else \expandafter \@secondoftwo
 \fi
}%
\providecommand \natexlab [1]{#1}%
\providecommand \enquote  [1]{``#1''}%
\providecommand \bibnamefont  [1]{#1}%
\providecommand \bibfnamefont [1]{#1}%
\providecommand \citenamefont [1]{#1}%
\providecommand \href@noop [0]{\@secondoftwo}%
\providecommand \href [0]{\begingroup \@sanitize@url \@href}%
\providecommand \@href[1]{\@@startlink{#1}\@@href}%
\providecommand \@@href[1]{\endgroup#1\@@endlink}%
\providecommand \@sanitize@url [0]{\catcode `\\12\catcode `\$12\catcode `\&12\catcode `\#12\catcode `\^12\catcode `\_12\catcode `\%12\relax}%
\providecommand \@@startlink[1]{}%
\providecommand \@@endlink[0]{}%
\providecommand \url  [0]{\begingroup\@sanitize@url \@url }%
\providecommand \@url [1]{\endgroup\@href {#1}{\urlprefix }}%
\providecommand \urlprefix  [0]{URL }%
\providecommand \Eprint [0]{\href }%
\providecommand \doibase [0]{http://dx.doi.org/}%
\providecommand \selectlanguage [0]{\@gobble}%
\providecommand \bibinfo  [0]{\@secondoftwo}%
\providecommand \bibfield  [0]{\@secondoftwo}%
\providecommand \translation [1]{[#1]}%
\providecommand \BibitemOpen [0]{}%
\providecommand \bibitemStop [0]{}%
\providecommand \bibitemNoStop [0]{.\EOS\space}%
\providecommand \EOS [0]{\spacefactor3000\relax}%
\providecommand \BibitemShut  [1]{\csname bibitem#1\endcsname}%
\let\auto@bib@innerbib\@empty
\bibitem [{\citenamefont {Harris}\ and\ \citenamefont {M\"uller}(2024)}]{Harris:2024aov}%
  \BibitemOpen
  \bibfield  {author} {\bibinfo {author} {\bibfnamefont {J.~W.}\ \bibnamefont {Harris}}\ and\ \bibinfo {author} {\bibfnamefont {B.}~\bibnamefont {M\"uller}},\ }\href {\doibase 10.1140/epjc/s10052-024-12533-y} {\bibfield  {journal} {\bibinfo  {journal} {Eur. Phys. J. C}\ }\textbf {\bibinfo {volume} {84}},\ \bibinfo {pages} {247} (\bibinfo {year} {2024})}\BibitemShut {NoStop}%
\bibitem [{\citenamefont {Heinz}\ and\ \citenamefont {Snellings}(2013)}]{Heinz:2013th}%
  \BibitemOpen
  \bibfield  {author} {\bibinfo {author} {\bibfnamefont {U.}~\bibnamefont {Heinz}}\ and\ \bibinfo {author} {\bibfnamefont {R.}~\bibnamefont {Snellings}},\ }\href {\doibase 10.1146/annurev-nucl-102212-170540} {\bibfield  {journal} {\bibinfo  {journal} {Ann. Rev. Nucl. Part. Sci.}\ }\textbf {\bibinfo {volume} {63}},\ \bibinfo {pages} {123} (\bibinfo {year} {2013})},\ \Eprint {http://arxiv.org/abs/1301.2826} {arXiv:1301.2826 [nucl-th]} \BibitemShut {NoStop}%
\bibitem [{\citenamefont {Ratti}\ and\ \citenamefont {Bellwied}(2021)}]{Ratti:2021ubw}%
  \BibitemOpen
  \bibfield  {author} {\bibinfo {author} {\bibfnamefont {C.}~\bibnamefont {Ratti}}\ and\ \bibinfo {author} {\bibfnamefont {R.}~\bibnamefont {Bellwied}},\ }\href {\doibase 10.1007/978-3-030-67235-5} {\emph {\bibinfo {title} {{The Deconfinement Transition of QCD: Theory Meets Experiment}}}},\ \bibinfo {series} {Lecture Notes in Physics}, Vol.\ \bibinfo {volume} {981}\ (\bibinfo {year} {2021})\BibitemShut {NoStop}%
\bibitem [{\citenamefont {Yagi}\ \emph {et~al.}(2005)\citenamefont {Yagi}, \citenamefont {Hatsuda},\ and\ \citenamefont {Miake}}]{Yagi:2005yb}%
  \BibitemOpen
  \bibfield  {author} {\bibinfo {author} {\bibfnamefont {K.}~\bibnamefont {Yagi}}, \bibinfo {author} {\bibfnamefont {T.}~\bibnamefont {Hatsuda}}, \ and\ \bibinfo {author} {\bibfnamefont {Y.}~\bibnamefont {Miake}},\ }\href@noop {} {\emph {\bibinfo {title} {{Quark-gluon plasma: From big bang to little bang}}}},\ Vol.~\bibinfo {volume} {23}\ (\bibinfo {year} {2005})\BibitemShut {NoStop}%
\bibitem [{\citenamefont {Gale}\ \emph {et~al.}(2013)\citenamefont {Gale}, \citenamefont {Jeon},\ and\ \citenamefont {Schenke}}]{Gale:2013da}%
  \BibitemOpen
  \bibfield  {author} {\bibinfo {author} {\bibfnamefont {C.}~\bibnamefont {Gale}}, \bibinfo {author} {\bibfnamefont {S.}~\bibnamefont {Jeon}}, \ and\ \bibinfo {author} {\bibfnamefont {B.}~\bibnamefont {Schenke}},\ }\href {\doibase 10.1142/S0217751X13400113} {\bibfield  {journal} {\bibinfo  {journal} {Int. J. Mod. Phys. A}\ }\textbf {\bibinfo {volume} {28}},\ \bibinfo {pages} {1340011} (\bibinfo {year} {2013})},\ \Eprint {http://arxiv.org/abs/1301.5893} {arXiv:1301.5893 [nucl-th]} \BibitemShut {NoStop}%
\bibitem [{\citenamefont {Busza}\ \emph {et~al.}(2018)\citenamefont {Busza}, \citenamefont {Rajagopal},\ and\ \citenamefont {van~der Schee}}]{Busza:2018rrf}%
  \BibitemOpen
  \bibfield  {author} {\bibinfo {author} {\bibfnamefont {W.}~\bibnamefont {Busza}}, \bibinfo {author} {\bibfnamefont {K.}~\bibnamefont {Rajagopal}}, \ and\ \bibinfo {author} {\bibfnamefont {W.}~\bibnamefont {van~der Schee}},\ }\href {\doibase 10.1146/annurev-nucl-101917-020852} {\bibfield  {journal} {\bibinfo  {journal} {Ann. Rev. Nucl. Part. Sci.}\ }\textbf {\bibinfo {volume} {68}},\ \bibinfo {pages} {339} (\bibinfo {year} {2018})},\ \Eprint {http://arxiv.org/abs/1802.04801} {arXiv:1802.04801 [hep-ph]} \BibitemShut {NoStop}%
\bibitem [{\citenamefont {Derradi~de Souza}\ \emph {et~al.}(2016)\citenamefont {Derradi~de Souza}, \citenamefont {Koide},\ and\ \citenamefont {Kodama}}]{DerradideSouza:2015kpt}%
  \BibitemOpen
  \bibfield  {author} {\bibinfo {author} {\bibfnamefont {R.}~\bibnamefont {Derradi~de Souza}}, \bibinfo {author} {\bibfnamefont {T.}~\bibnamefont {Koide}}, \ and\ \bibinfo {author} {\bibfnamefont {T.}~\bibnamefont {Kodama}},\ }\href {\doibase 10.1016/j.ppnp.2015.09.002} {\bibfield  {journal} {\bibinfo  {journal} {Prog. Part. Nucl. Phys.}\ }\textbf {\bibinfo {volume} {86}},\ \bibinfo {pages} {35} (\bibinfo {year} {2016})},\ \Eprint {http://arxiv.org/abs/1506.03863} {arXiv:1506.03863 [nucl-th]} \BibitemShut {NoStop}%
\bibitem [{\citenamefont {Niida}\ and\ \citenamefont {Miake}(2021)}]{Niida:2021wut}%
  \BibitemOpen
  \bibfield  {author} {\bibinfo {author} {\bibfnamefont {T.}~\bibnamefont {Niida}}\ and\ \bibinfo {author} {\bibfnamefont {Y.}~\bibnamefont {Miake}},\ }\href {\doibase 10.1007/s43673-021-00014-3} {\bibfield  {journal} {\bibinfo  {journal} {AAPPS Bull.}\ }\textbf {\bibinfo {volume} {31}},\ \bibinfo {pages} {12} (\bibinfo {year} {2021})},\ \Eprint {http://arxiv.org/abs/2104.11406} {arXiv:2104.11406 [nucl-ex]} \BibitemShut {NoStop}%
\bibitem [{\citenamefont {Bazavov}\ \emph {et~al.}(2009)\citenamefont {Bazavov} \emph {et~al.}}]{Bazavov:2009zn}%
  \BibitemOpen
  \bibfield  {author} {\bibinfo {author} {\bibfnamefont {A.}~\bibnamefont {Bazavov}} \emph {et~al.},\ }\href {\doibase 10.1103/PhysRevD.80.014504} {\bibfield  {journal} {\bibinfo  {journal} {Phys. Rev. D}\ }\textbf {\bibinfo {volume} {80}},\ \bibinfo {pages} {014504} (\bibinfo {year} {2009})},\ \Eprint {http://arxiv.org/abs/0903.4379} {arXiv:0903.4379 [hep-lat]} \BibitemShut {NoStop}%
\bibitem [{\citenamefont {Borsanyi}\ \emph {et~al.}(2010)\citenamefont {Borsanyi}, \citenamefont {Endrodi}, \citenamefont {Fodor}, \citenamefont {Jakovac}, \citenamefont {Katz}, \citenamefont {Krieg}, \citenamefont {Ratti},\ and\ \citenamefont {Szabo}}]{Borsanyi:2010cj}%
  \BibitemOpen
  \bibfield  {author} {\bibinfo {author} {\bibfnamefont {S.}~\bibnamefont {Borsanyi}}, \bibinfo {author} {\bibfnamefont {G.}~\bibnamefont {Endrodi}}, \bibinfo {author} {\bibfnamefont {Z.}~\bibnamefont {Fodor}}, \bibinfo {author} {\bibfnamefont {A.}~\bibnamefont {Jakovac}}, \bibinfo {author} {\bibfnamefont {S.~D.}\ \bibnamefont {Katz}}, \bibinfo {author} {\bibfnamefont {S.}~\bibnamefont {Krieg}}, \bibinfo {author} {\bibfnamefont {C.}~\bibnamefont {Ratti}}, \ and\ \bibinfo {author} {\bibfnamefont {K.~K.}\ \bibnamefont {Szabo}},\ }\href {\doibase 10.1007/JHEP11(2010)077} {\bibfield  {journal} {\bibinfo  {journal} {JHEP}\ }\textbf {\bibinfo {volume} {11}},\ \bibinfo {pages} {077} (\bibinfo {year} {2010})},\ \Eprint {http://arxiv.org/abs/1007.2580} {arXiv:1007.2580 [hep-lat]} \BibitemShut {NoStop}%
\bibitem [{\citenamefont {Bazavov}\ \emph {et~al.}(2017)\citenamefont {Bazavov} \emph {et~al.}}]{Bazavov:2017dus}%
  \BibitemOpen
  \bibfield  {author} {\bibinfo {author} {\bibfnamefont {A.}~\bibnamefont {Bazavov}} \emph {et~al.},\ }\href {\doibase 10.1103/PhysRevD.95.054504} {\bibfield  {journal} {\bibinfo  {journal} {Phys. Rev. D}\ }\textbf {\bibinfo {volume} {95}},\ \bibinfo {pages} {054504} (\bibinfo {year} {2017})},\ \Eprint {http://arxiv.org/abs/1701.04325} {arXiv:1701.04325 [hep-lat]} \BibitemShut {NoStop}%
\bibitem [{\citenamefont {Bors\'anyi}\ \emph {et~al.}(2021)\citenamefont {Bors\'anyi}, \citenamefont {Fodor}, \citenamefont {Guenther}, \citenamefont {Kara}, \citenamefont {Katz}, \citenamefont {Parotto}, \citenamefont {P\'asztor}, \citenamefont {Ratti},\ and\ \citenamefont {Szab\'o}}]{Borsanyi:2021sxv}%
  \BibitemOpen
  \bibfield  {author} {\bibinfo {author} {\bibfnamefont {S.}~\bibnamefont {Bors\'anyi}}, \bibinfo {author} {\bibfnamefont {Z.}~\bibnamefont {Fodor}}, \bibinfo {author} {\bibfnamefont {J.~N.}\ \bibnamefont {Guenther}}, \bibinfo {author} {\bibfnamefont {R.}~\bibnamefont {Kara}}, \bibinfo {author} {\bibfnamefont {S.~D.}\ \bibnamefont {Katz}}, \bibinfo {author} {\bibfnamefont {P.}~\bibnamefont {Parotto}}, \bibinfo {author} {\bibfnamefont {A.}~\bibnamefont {P\'asztor}}, \bibinfo {author} {\bibfnamefont {C.}~\bibnamefont {Ratti}}, \ and\ \bibinfo {author} {\bibfnamefont {K.~K.}\ \bibnamefont {Szab\'o}},\ }\href {\doibase 10.1103/PhysRevLett.126.232001} {\bibfield  {journal} {\bibinfo  {journal} {Phys. Rev. Lett.}\ }\textbf {\bibinfo {volume} {126}},\ \bibinfo {pages} {232001} (\bibinfo {year} {2021})},\ \Eprint {http://arxiv.org/abs/2102.06660} {arXiv:2102.06660 [hep-lat]} \BibitemShut {NoStop}%
\bibitem [{\citenamefont {Ratti}(2018)}]{Ratti:2018ksb}%
  \BibitemOpen
  \bibfield  {author} {\bibinfo {author} {\bibfnamefont {C.}~\bibnamefont {Ratti}},\ }\href {\doibase 10.1088/1361-6633/aabb97} {\bibfield  {journal} {\bibinfo  {journal} {Rept. Prog. Phys.}\ }\textbf {\bibinfo {volume} {81}},\ \bibinfo {pages} {084301} (\bibinfo {year} {2018})},\ \Eprint {http://arxiv.org/abs/1804.07810} {arXiv:1804.07810 [hep-lat]} \BibitemShut {NoStop}%
\bibitem [{\citenamefont {Venugopalan}\ and\ \citenamefont {Prakash}(1992)}]{Venugopalan:1992hy}%
  \BibitemOpen
  \bibfield  {author} {\bibinfo {author} {\bibfnamefont {R.}~\bibnamefont {Venugopalan}}\ and\ \bibinfo {author} {\bibfnamefont {M.}~\bibnamefont {Prakash}},\ }\href {\doibase 10.1016/0375-9474(92)90005-5} {\bibfield  {journal} {\bibinfo  {journal} {Nucl. Phys. A}\ }\textbf {\bibinfo {volume} {546}},\ \bibinfo {pages} {718} (\bibinfo {year} {1992})}\BibitemShut {NoStop}%
\bibitem [{\citenamefont {Karsch}\ \emph {et~al.}(2003)\citenamefont {Karsch}, \citenamefont {Redlich},\ and\ \citenamefont {Tawfik}}]{Karsch:2003vd}%
  \BibitemOpen
  \bibfield  {author} {\bibinfo {author} {\bibfnamefont {F.}~\bibnamefont {Karsch}}, \bibinfo {author} {\bibfnamefont {K.}~\bibnamefont {Redlich}}, \ and\ \bibinfo {author} {\bibfnamefont {A.}~\bibnamefont {Tawfik}},\ }\href {\doibase 10.1140/epjc/s2003-01228-y} {\bibfield  {journal} {\bibinfo  {journal} {Eur. Phys. J. C}\ }\textbf {\bibinfo {volume} {29}},\ \bibinfo {pages} {549} (\bibinfo {year} {2003})},\ \Eprint {http://arxiv.org/abs/hep-ph/0303108} {arXiv:hep-ph/0303108} \BibitemShut {NoStop}%
\bibitem [{\citenamefont {Huovinen}\ and\ \citenamefont {Petreczky}(2010)}]{Huovinen:2009yb}%
  \BibitemOpen
  \bibfield  {author} {\bibinfo {author} {\bibfnamefont {P.}~\bibnamefont {Huovinen}}\ and\ \bibinfo {author} {\bibfnamefont {P.}~\bibnamefont {Petreczky}},\ }\href {\doibase 10.1016/j.nuclphysa.2010.02.015} {\bibfield  {journal} {\bibinfo  {journal} {Nucl. Phys. A}\ }\textbf {\bibinfo {volume} {837}},\ \bibinfo {pages} {26} (\bibinfo {year} {2010})},\ \Eprint {http://arxiv.org/abs/0912.2541} {arXiv:0912.2541 [hep-ph]} \BibitemShut {NoStop}%
\bibitem [{\citenamefont {Sorensen}\ \emph {et~al.}(2024)\citenamefont {Sorensen} \emph {et~al.}}]{Sorensen:2023zkk}%
  \BibitemOpen
  \bibfield  {author} {\bibinfo {author} {\bibfnamefont {A.}~\bibnamefont {Sorensen}} \emph {et~al.},\ }\href {\doibase 10.1016/j.ppnp.2023.104080} {\bibfield  {journal} {\bibinfo  {journal} {Prog. Part. Nucl. Phys.}\ }\textbf {\bibinfo {volume} {134}},\ \bibinfo {pages} {104080} (\bibinfo {year} {2024})},\ \Eprint {http://arxiv.org/abs/2301.13253} {arXiv:2301.13253 [nucl-th]} \BibitemShut {NoStop}%
\bibitem [{\citenamefont {Kumar}\ \emph {et~al.}(2024)\citenamefont {Kumar} \emph {et~al.}}]{MUSES:2023hyz}%
  \BibitemOpen
  \bibfield  {author} {\bibinfo {author} {\bibfnamefont {R.}~\bibnamefont {Kumar}} \emph {et~al.} (\bibinfo {collaboration} {MUSES}),\ }\href {\doibase 10.1007/s41114-024-00049-6} {\bibfield  {journal} {\bibinfo  {journal} {Living Rev. Rel.}\ }\textbf {\bibinfo {volume} {27}},\ \bibinfo {pages} {3} (\bibinfo {year} {2024})},\ \Eprint {http://arxiv.org/abs/2303.17021} {arXiv:2303.17021 [nucl-th]} \BibitemShut {NoStop}%
\bibitem [{\citenamefont {Du}\ \emph {et~al.}(2024)\citenamefont {Du}, \citenamefont {Sorensen},\ and\ \citenamefont {Stephanov}}]{Du:2024wjm}%
  \BibitemOpen
  \bibfield  {author} {\bibinfo {author} {\bibfnamefont {L.}~\bibnamefont {Du}}, \bibinfo {author} {\bibfnamefont {A.}~\bibnamefont {Sorensen}}, \ and\ \bibinfo {author} {\bibfnamefont {M.}~\bibnamefont {Stephanov}}\ }(\bibinfo {year} {2024})\ \Eprint {http://arxiv.org/abs/2402.10183} {arXiv:2402.10183 [nucl-th]} \BibitemShut {NoStop}%
\bibitem [{\citenamefont {Auvinen}\ \emph {et~al.}(2020)\citenamefont {Auvinen}, \citenamefont {Eskola}, \citenamefont {Huovinen}, \citenamefont {Niemi}, \citenamefont {Paatelainen},\ and\ \citenamefont {Petreczky}}]{Auvinen:2020mpc}%
  \BibitemOpen
  \bibfield  {author} {\bibinfo {author} {\bibfnamefont {J.}~\bibnamefont {Auvinen}}, \bibinfo {author} {\bibfnamefont {K.~J.}\ \bibnamefont {Eskola}}, \bibinfo {author} {\bibfnamefont {P.}~\bibnamefont {Huovinen}}, \bibinfo {author} {\bibfnamefont {H.}~\bibnamefont {Niemi}}, \bibinfo {author} {\bibfnamefont {R.}~\bibnamefont {Paatelainen}}, \ and\ \bibinfo {author} {\bibfnamefont {P.}~\bibnamefont {Petreczky}},\ }\href {\doibase 10.1103/PhysRevC.102.044911} {\bibfield  {journal} {\bibinfo  {journal} {Phys. Rev. C}\ }\textbf {\bibinfo {volume} {102}},\ \bibinfo {pages} {044911} (\bibinfo {year} {2020})},\ \Eprint {http://arxiv.org/abs/2006.12499} {arXiv:2006.12499 [nucl-th]} \BibitemShut {NoStop}%
\bibitem [{\citenamefont {Moreland}\ and\ \citenamefont {Soltz}(2016)}]{Moreland:2015dvc}%
  \BibitemOpen
  \bibfield  {author} {\bibinfo {author} {\bibfnamefont {J.~S.}\ \bibnamefont {Moreland}}\ and\ \bibinfo {author} {\bibfnamefont {R.~A.}\ \bibnamefont {Soltz}},\ }\href {\doibase 10.1103/PhysRevC.93.044913} {\bibfield  {journal} {\bibinfo  {journal} {Phys. Rev. C}\ }\textbf {\bibinfo {volume} {93}},\ \bibinfo {pages} {044913} (\bibinfo {year} {2016})},\ \Eprint {http://arxiv.org/abs/1512.02189} {arXiv:1512.02189 [nucl-th]} \BibitemShut {NoStop}%
\bibitem [{\citenamefont {Pratt}\ \emph {et~al.}(2015)\citenamefont {Pratt}, \citenamefont {Sangaline}, \citenamefont {Sorensen},\ and\ \citenamefont {Wang}}]{Pratt:2015zsa}%
  \BibitemOpen
  \bibfield  {author} {\bibinfo {author} {\bibfnamefont {S.}~\bibnamefont {Pratt}}, \bibinfo {author} {\bibfnamefont {E.}~\bibnamefont {Sangaline}}, \bibinfo {author} {\bibfnamefont {P.}~\bibnamefont {Sorensen}}, \ and\ \bibinfo {author} {\bibfnamefont {H.}~\bibnamefont {Wang}},\ }\href {\doibase 10.1103/PhysRevLett.114.202301} {\bibfield  {journal} {\bibinfo  {journal} {Phys. Rev. Lett.}\ }\textbf {\bibinfo {volume} {114}},\ \bibinfo {pages} {202301} (\bibinfo {year} {2015})},\ \Eprint {http://arxiv.org/abs/1501.04042} {arXiv:1501.04042 [nucl-th]} \BibitemShut {NoStop}%
\bibitem [{\citenamefont {Blaizot}\ and\ \citenamefont {Ollitrault}(1987)}]{Blaizot:1987cc}%
  \BibitemOpen
  \bibfield  {author} {\bibinfo {author} {\bibfnamefont {J.~P.}\ \bibnamefont {Blaizot}}\ and\ \bibinfo {author} {\bibfnamefont {J.-Y.}\ \bibnamefont {Ollitrault}},\ }\href {\doibase 10.1016/0370-2693(87)91314-1} {\bibfield  {journal} {\bibinfo  {journal} {Phys. Lett. B}\ }\textbf {\bibinfo {volume} {191}},\ \bibinfo {pages} {21} (\bibinfo {year} {1987})}\BibitemShut {NoStop}%
\bibitem [{\citenamefont {Campanini}\ and\ \citenamefont {Ferri}(2011)}]{Campanini:2011bj}%
  \BibitemOpen
  \bibfield  {author} {\bibinfo {author} {\bibfnamefont {R.}~\bibnamefont {Campanini}}\ and\ \bibinfo {author} {\bibfnamefont {G.}~\bibnamefont {Ferri}},\ }\href {\doibase 10.1016/j.physletb.2011.08.009} {\bibfield  {journal} {\bibinfo  {journal} {Phys. Lett. B}\ }\textbf {\bibinfo {volume} {703}},\ \bibinfo {pages} {237} (\bibinfo {year} {2011})},\ \Eprint {http://arxiv.org/abs/1106.2008} {arXiv:1106.2008 [hep-ph]} \BibitemShut {NoStop}%
\bibitem [{\citenamefont {Monnai}\ and\ \citenamefont {Ollitrault}(2017)}]{Monnai:2017cbv}%
  \BibitemOpen
  \bibfield  {author} {\bibinfo {author} {\bibfnamefont {A.}~\bibnamefont {Monnai}}\ and\ \bibinfo {author} {\bibfnamefont {J.-Y.}\ \bibnamefont {Ollitrault}},\ }\href {\doibase 10.1103/PhysRevC.96.044902} {\bibfield  {journal} {\bibinfo  {journal} {Phys. Rev. C}\ }\textbf {\bibinfo {volume} {96}},\ \bibinfo {pages} {044902} (\bibinfo {year} {2017})},\ \Eprint {http://arxiv.org/abs/1707.08466} {arXiv:1707.08466 [nucl-th]} \BibitemShut {NoStop}%
\bibitem [{\citenamefont {Gardim}\ \emph {et~al.}(2020{\natexlab{a}})\citenamefont {Gardim}, \citenamefont {Giacalone}, \citenamefont {Luzum},\ and\ \citenamefont {Ollitrault}}]{Gardim:2019xjs}%
  \BibitemOpen
  \bibfield  {author} {\bibinfo {author} {\bibfnamefont {F.~G.}\ \bibnamefont {Gardim}}, \bibinfo {author} {\bibfnamefont {G.}~\bibnamefont {Giacalone}}, \bibinfo {author} {\bibfnamefont {M.}~\bibnamefont {Luzum}}, \ and\ \bibinfo {author} {\bibfnamefont {J.-Y.}\ \bibnamefont {Ollitrault}},\ }\href {\doibase 10.1038/s41567-020-0846-4} {\bibfield  {journal} {\bibinfo  {journal} {Nature Phys.}\ }\textbf {\bibinfo {volume} {16}},\ \bibinfo {pages} {615} (\bibinfo {year} {2020}{\natexlab{a}})},\ \Eprint {http://arxiv.org/abs/1908.09728} {arXiv:1908.09728 [nucl-th]} \BibitemShut {NoStop}%
\bibitem [{\citenamefont {Gardim}\ \emph {et~al.}(2020{\natexlab{b}})\citenamefont {Gardim}, \citenamefont {Giacalone},\ and\ \citenamefont {Ollitrault}}]{Gardim:2019brr}%
  \BibitemOpen
  \bibfield  {author} {\bibinfo {author} {\bibfnamefont {F.~G.}\ \bibnamefont {Gardim}}, \bibinfo {author} {\bibfnamefont {G.}~\bibnamefont {Giacalone}}, \ and\ \bibinfo {author} {\bibfnamefont {J.-Y.}\ \bibnamefont {Ollitrault}},\ }\href {\doibase 10.1016/j.physletb.2020.135749} {\bibfield  {journal} {\bibinfo  {journal} {Phys. Lett. B}\ }\textbf {\bibinfo {volume} {809}},\ \bibinfo {pages} {135749} (\bibinfo {year} {2020}{\natexlab{b}})},\ \Eprint {http://arxiv.org/abs/1909.11609} {arXiv:1909.11609 [nucl-th]} \BibitemShut {NoStop}%
\bibitem [{\citenamefont {Van~Hove}(1982)}]{VanHove:1982vk}%
  \BibitemOpen
  \bibfield  {author} {\bibinfo {author} {\bibfnamefont {L.}~\bibnamefont {Van~Hove}},\ }\href {\doibase 10.1016/0370-2693(82)90617-7} {\bibfield  {journal} {\bibinfo  {journal} {Phys. Lett. B}\ }\textbf {\bibinfo {volume} {118}},\ \bibinfo {pages} {138} (\bibinfo {year} {1982})}\BibitemShut {NoStop}%
\bibitem [{\citenamefont {Hayrapetyan}\ \emph {et~al.}(2024)\citenamefont {Hayrapetyan} \emph {et~al.}}]{CMS:2024sgx}%
  \BibitemOpen
  \bibfield  {author} {\bibinfo {author} {\bibfnamefont {A.}~\bibnamefont {Hayrapetyan}} \emph {et~al.} (\bibinfo {collaboration} {CMS}),\ }\href {\doibase 10.1088/1361-6633/ad4b9b} {\bibfield  {journal} {\bibinfo  {journal} {Rept. Prog. Phys.}\ }\textbf {\bibinfo {volume} {87}},\ \bibinfo {pages} {077801} (\bibinfo {year} {2024})},\ \Eprint {http://arxiv.org/abs/2401.06896} {arXiv:2401.06896 [nucl-ex]} \BibitemShut {NoStop}%
\bibitem [{\citenamefont {Bazavov}\ \emph {et~al.}(2014)\citenamefont {Bazavov} \emph {et~al.}}]{HotQCD:2014kol}%
  \BibitemOpen
  \bibfield  {author} {\bibinfo {author} {\bibfnamefont {A.}~\bibnamefont {Bazavov}} \emph {et~al.} (\bibinfo {collaboration} {HotQCD}),\ }\href {\doibase 10.1103/PhysRevD.90.094503} {\bibfield  {journal} {\bibinfo  {journal} {Phys. Rev. D}\ }\textbf {\bibinfo {volume} {90}},\ \bibinfo {pages} {094503} (\bibinfo {year} {2014})},\ \Eprint {http://arxiv.org/abs/1407.6387} {arXiv:1407.6387 [hep-lat]} \BibitemShut {NoStop}%
\bibitem [{\citenamefont {Nijs}\ and\ \citenamefont {van~der Schee}(2024)}]{Nijs:2023bzv}%
  \BibitemOpen
  \bibfield  {author} {\bibinfo {author} {\bibfnamefont {G.}~\bibnamefont {Nijs}}\ and\ \bibinfo {author} {\bibfnamefont {W.}~\bibnamefont {van~der Schee}},\ }\href {\doibase 10.1016/j.physletb.2024.138636} {\bibfield  {journal} {\bibinfo  {journal} {Phys. Lett. B}\ }\textbf {\bibinfo {volume} {853}},\ \bibinfo {pages} {138636} (\bibinfo {year} {2024})},\ \Eprint {http://arxiv.org/abs/2312.04623} {arXiv:2312.04623 [nucl-th]} \BibitemShut {NoStop}%
\bibitem [{\citenamefont {Gardim}\ \emph {et~al.}(2024)\citenamefont {Gardim}, \citenamefont {Giannini},\ and\ \citenamefont {Ollitrault}}]{Gardim:2024zvi}%
  \BibitemOpen
  \bibfield  {author} {\bibinfo {author} {\bibfnamefont {F.~G.}\ \bibnamefont {Gardim}}, \bibinfo {author} {\bibfnamefont {A.~V.}\ \bibnamefont {Giannini}}, \ and\ \bibinfo {author} {\bibfnamefont {J.-Y.}\ \bibnamefont {Ollitrault}},\ }\href {\doibase 10.1016/j.physletb.2024.138937} {\bibfield  {journal} {\bibinfo  {journal} {Phys. Lett. B}\ }\textbf {\bibinfo {volume} {856}},\ \bibinfo {pages} {138937} (\bibinfo {year} {2024})},\ \Eprint {http://arxiv.org/abs/2403.06052} {arXiv:2403.06052 [nucl-th]} \BibitemShut {NoStop}%
\bibitem [{\citenamefont {Baym}\ \emph {et~al.}(1983)\citenamefont {Baym}, \citenamefont {Friman}, \citenamefont {Blaizot}, \citenamefont {Soyeur},\ and\ \citenamefont {Czyz}}]{Baym:1983amj}%
  \BibitemOpen
  \bibfield  {author} {\bibinfo {author} {\bibfnamefont {G.}~\bibnamefont {Baym}}, \bibinfo {author} {\bibfnamefont {B.~L.}\ \bibnamefont {Friman}}, \bibinfo {author} {\bibfnamefont {J.~P.}\ \bibnamefont {Blaizot}}, \bibinfo {author} {\bibfnamefont {M.}~\bibnamefont {Soyeur}}, \ and\ \bibinfo {author} {\bibfnamefont {W.}~\bibnamefont {Czyz}},\ }\href {\doibase 10.1016/0375-9474(83)90666-8} {\bibfield  {journal} {\bibinfo  {journal} {Nucl. Phys. A}\ }\textbf {\bibinfo {volume} {407}},\ \bibinfo {pages} {541} (\bibinfo {year} {1983})}\BibitemShut {NoStop}%
\bibitem [{\citenamefont {Ruuskanen}(1987)}]{Ruuskanen:1986py}%
  \BibitemOpen
  \bibfield  {author} {\bibinfo {author} {\bibfnamefont {P.~V.}\ \bibnamefont {Ruuskanen}},\ }\href@noop {} {\bibfield  {journal} {\bibinfo  {journal} {Acta Phys. Polon. B}\ }\textbf {\bibinfo {volume} {18}},\ \bibinfo {pages} {551} (\bibinfo {year} {1987})}\BibitemShut {NoStop}%
\bibitem [{\citenamefont {Gubser}\ and\ \citenamefont {Yarom}(2011)}]{Gubser:2010ui}%
  \BibitemOpen
  \bibfield  {author} {\bibinfo {author} {\bibfnamefont {S.~S.}\ \bibnamefont {Gubser}}\ and\ \bibinfo {author} {\bibfnamefont {A.}~\bibnamefont {Yarom}},\ }\href {\doibase 10.1016/j.nuclphysb.2011.01.012} {\bibfield  {journal} {\bibinfo  {journal} {Nucl. Phys. B}\ }\textbf {\bibinfo {volume} {846}},\ \bibinfo {pages} {469} (\bibinfo {year} {2011})},\ \Eprint {http://arxiv.org/abs/1012.1314} {arXiv:1012.1314 [hep-th]} \BibitemShut {NoStop}%
\bibitem [{\citenamefont {Gubser}(2010)}]{Gubser:2010ze}%
  \BibitemOpen
  \bibfield  {author} {\bibinfo {author} {\bibfnamefont {S.~S.}\ \bibnamefont {Gubser}},\ }\href {\doibase 10.1103/PhysRevD.82.085027} {\bibfield  {journal} {\bibinfo  {journal} {Phys. Rev. D}\ }\textbf {\bibinfo {volume} {82}},\ \bibinfo {pages} {085027} (\bibinfo {year} {2010})},\ \Eprint {http://arxiv.org/abs/1006.0006} {arXiv:1006.0006 [hep-th]} \BibitemShut {NoStop}%
\bibitem [{\citenamefont {Landau}\ and\ \citenamefont {Lifshitz}(2013)}]{landau2013fluid}%
  \BibitemOpen
  \bibfield  {author} {\bibinfo {author} {\bibfnamefont {L.~D.}\ \bibnamefont {Landau}}\ and\ \bibinfo {author} {\bibfnamefont {E.~M.}\ \bibnamefont {Lifshitz}},\ }\href@noop {} {\emph {\bibinfo {title} {Fluid mechanics: Landau And Lifshitz: course of theoretical physics, Volume 6}}},\ Vol.~\bibinfo {volume} {6}\ (\bibinfo  {publisher} {Elsevier},\ \bibinfo {year} {2013})\BibitemShut {NoStop}%
\bibitem [{\citenamefont {Cooper}\ and\ \citenamefont {Frye}(1974)}]{Cooper:1974mv}%
  \BibitemOpen
  \bibfield  {author} {\bibinfo {author} {\bibfnamefont {F.}~\bibnamefont {Cooper}}\ and\ \bibinfo {author} {\bibfnamefont {G.}~\bibnamefont {Frye}},\ }\href {\doibase 10.1103/PhysRevD.10.186} {\bibfield  {journal} {\bibinfo  {journal} {Phys. Rev. D}\ }\textbf {\bibinfo {volume} {10}},\ \bibinfo {pages} {186} (\bibinfo {year} {1974})}\BibitemShut {NoStop}%
\bibitem [{\citenamefont {Schenke}\ \emph {et~al.}(2010)\citenamefont {Schenke}, \citenamefont {Jeon},\ and\ \citenamefont {Gale}}]{Schenke:2010nt}%
  \BibitemOpen
  \bibfield  {author} {\bibinfo {author} {\bibfnamefont {B.}~\bibnamefont {Schenke}}, \bibinfo {author} {\bibfnamefont {S.}~\bibnamefont {Jeon}}, \ and\ \bibinfo {author} {\bibfnamefont {C.}~\bibnamefont {Gale}},\ }\href {\doibase 10.1103/PhysRevC.82.014903} {\bibfield  {journal} {\bibinfo  {journal} {Phys. Rev. C}\ }\textbf {\bibinfo {volume} {82}},\ \bibinfo {pages} {014903} (\bibinfo {year} {2010})},\ \Eprint {http://arxiv.org/abs/1004.1408} {arXiv:1004.1408 [hep-ph]} \BibitemShut {NoStop}%
\bibitem [{\citenamefont {Ryu}\ \emph {et~al.}(2015)\citenamefont {Ryu}, \citenamefont {Paquet}, \citenamefont {Shen}, \citenamefont {Denicol}, \citenamefont {Schenke}, \citenamefont {Jeon},\ and\ \citenamefont {Gale}}]{Ryu:2015vwa}%
  \BibitemOpen
  \bibfield  {author} {\bibinfo {author} {\bibfnamefont {S.}~\bibnamefont {Ryu}}, \bibinfo {author} {\bibfnamefont {J.~F.}\ \bibnamefont {Paquet}}, \bibinfo {author} {\bibfnamefont {C.}~\bibnamefont {Shen}}, \bibinfo {author} {\bibfnamefont {G.~S.}\ \bibnamefont {Denicol}}, \bibinfo {author} {\bibfnamefont {B.}~\bibnamefont {Schenke}}, \bibinfo {author} {\bibfnamefont {S.}~\bibnamefont {Jeon}}, \ and\ \bibinfo {author} {\bibfnamefont {C.}~\bibnamefont {Gale}},\ }\href {\doibase 10.1103/PhysRevLett.115.132301} {\bibfield  {journal} {\bibinfo  {journal} {Phys. Rev. Lett.}\ }\textbf {\bibinfo {volume} {115}},\ \bibinfo {pages} {132301} (\bibinfo {year} {2015})},\ \Eprint {http://arxiv.org/abs/1502.01675} {arXiv:1502.01675 [nucl-th]} \BibitemShut {NoStop}%
\bibitem [{\citenamefont {Paquet}\ \emph {et~al.}(2016)\citenamefont {Paquet}, \citenamefont {Shen}, \citenamefont {Denicol}, \citenamefont {Luzum}, \citenamefont {Schenke}, \citenamefont {Jeon},\ and\ \citenamefont {Gale}}]{Paquet:2015lta}%
  \BibitemOpen
  \bibfield  {author} {\bibinfo {author} {\bibfnamefont {J.-F.}\ \bibnamefont {Paquet}}, \bibinfo {author} {\bibfnamefont {C.}~\bibnamefont {Shen}}, \bibinfo {author} {\bibfnamefont {G.~S.}\ \bibnamefont {Denicol}}, \bibinfo {author} {\bibfnamefont {M.}~\bibnamefont {Luzum}}, \bibinfo {author} {\bibfnamefont {B.}~\bibnamefont {Schenke}}, \bibinfo {author} {\bibfnamefont {S.}~\bibnamefont {Jeon}}, \ and\ \bibinfo {author} {\bibfnamefont {C.}~\bibnamefont {Gale}},\ }\href {\doibase 10.1103/PhysRevC.93.044906} {\bibfield  {journal} {\bibinfo  {journal} {Phys. Rev. C}\ }\textbf {\bibinfo {volume} {93}},\ \bibinfo {pages} {044906} (\bibinfo {year} {2016})},\ \Eprint {http://arxiv.org/abs/1509.06738} {arXiv:1509.06738 [hep-ph]} \BibitemShut {NoStop}%
\bibitem [{\citenamefont {Ryu}\ \emph {et~al.}(2018)\citenamefont {Ryu}, \citenamefont {Paquet}, \citenamefont {Shen}, \citenamefont {Denicol}, \citenamefont {Schenke}, \citenamefont {Jeon},\ and\ \citenamefont {Gale}}]{Ryu:2017qzn}%
  \BibitemOpen
  \bibfield  {author} {\bibinfo {author} {\bibfnamefont {S.}~\bibnamefont {Ryu}}, \bibinfo {author} {\bibfnamefont {J.-F.}\ \bibnamefont {Paquet}}, \bibinfo {author} {\bibfnamefont {C.}~\bibnamefont {Shen}}, \bibinfo {author} {\bibfnamefont {G.}~\bibnamefont {Denicol}}, \bibinfo {author} {\bibfnamefont {B.}~\bibnamefont {Schenke}}, \bibinfo {author} {\bibfnamefont {S.}~\bibnamefont {Jeon}}, \ and\ \bibinfo {author} {\bibfnamefont {C.}~\bibnamefont {Gale}},\ }\href {\doibase 10.1103/PhysRevC.97.034910} {\bibfield  {journal} {\bibinfo  {journal} {Phys. Rev. C}\ }\textbf {\bibinfo {volume} {97}},\ \bibinfo {pages} {034910} (\bibinfo {year} {2018})},\ \Eprint {http://arxiv.org/abs/1704.04216} {arXiv:1704.04216 [nucl-th]} \BibitemShut {NoStop}%
\bibitem [{\citenamefont {Hirano}(2002)}]{Hirano:2001eu}%
  \BibitemOpen
  \bibfield  {author} {\bibinfo {author} {\bibfnamefont {T.}~\bibnamefont {Hirano}},\ }\href {\doibase 10.1103/PhysRevC.65.011901} {\bibfield  {journal} {\bibinfo  {journal} {Phys. Rev. C}\ }\textbf {\bibinfo {volume} {65}},\ \bibinfo {pages} {011901} (\bibinfo {year} {2002})},\ \Eprint {http://arxiv.org/abs/nucl-th/0108004} {arXiv:nucl-th/0108004} \BibitemShut {NoStop}%
\bibitem [{\citenamefont {Cooper}\ \emph {et~al.}(1975)\citenamefont {Cooper}, \citenamefont {Frye},\ and\ \citenamefont {Schonberg}}]{Cooper:1974qi}%
  \BibitemOpen
  \bibfield  {author} {\bibinfo {author} {\bibfnamefont {F.}~\bibnamefont {Cooper}}, \bibinfo {author} {\bibfnamefont {G.}~\bibnamefont {Frye}}, \ and\ \bibinfo {author} {\bibfnamefont {E.}~\bibnamefont {Schonberg}},\ }\href {\doibase 10.1103/PhysRevD.11.192} {\bibfield  {journal} {\bibinfo  {journal} {Phys. Rev. D}\ }\textbf {\bibinfo {volume} {11}},\ \bibinfo {pages} {192} (\bibinfo {year} {1975})}\BibitemShut {NoStop}%
\bibitem [{\citenamefont {Vogt}(2007)}]{Vogt:2007zz}%
  \BibitemOpen
  \bibfield  {author} {\bibinfo {author} {\bibfnamefont {R.}~\bibnamefont {Vogt}},\ }\href@noop {} {\emph {\bibinfo {title} {{Ultrarelativistic heavy-ion collisions}}}}\ (\bibinfo  {publisher} {Elsevier},\ \bibinfo {address} {Amsterdam},\ \bibinfo {year} {2007})\BibitemShut {NoStop}%
\bibitem [{\citenamefont {Gradshteyn}\ and\ \citenamefont {Ryzhik}(2014)}]{gradshteyn2014table}%
  \BibitemOpen
  \bibfield  {author} {\bibinfo {author} {\bibfnamefont {I.}~\bibnamefont {Gradshteyn}}\ and\ \bibinfo {author} {\bibfnamefont {I.}~\bibnamefont {Ryzhik}},\ }\href@noop {} {\emph {\bibinfo {title} {Table of integrals, series, and products}}}\ (\bibinfo  {publisher} {Academic press},\ \bibinfo {year} {2014})\BibitemShut {NoStop}%
\bibitem [{\citenamefont {Denicol}\ and\ \citenamefont {Rischke}(2021)}]{Denicol:2021}%
  \BibitemOpen
  \bibfield  {author} {\bibinfo {author} {\bibfnamefont {G.~S.}\ \bibnamefont {Denicol}}\ and\ \bibinfo {author} {\bibfnamefont {D.~H.}\ \bibnamefont {Rischke}},\ }\href@noop {} {\emph {\bibinfo {title} {Microscopic Foundations of Relativistic Fluid Dynamics}}}\ (\bibinfo  {publisher} {Springer},\ \bibinfo {year} {2021})\BibitemShut {NoStop}%
\bibitem [{\citenamefont {Hatta}\ \emph {et~al.}(2014)\citenamefont {Hatta}, \citenamefont {Noronha}, \citenamefont {Torrieri},\ and\ \citenamefont {Xiao}}]{Hatta:2014jva}%
  \BibitemOpen
  \bibfield  {author} {\bibinfo {author} {\bibfnamefont {Y.}~\bibnamefont {Hatta}}, \bibinfo {author} {\bibfnamefont {J.}~\bibnamefont {Noronha}}, \bibinfo {author} {\bibfnamefont {G.}~\bibnamefont {Torrieri}}, \ and\ \bibinfo {author} {\bibfnamefont {B.-W.}\ \bibnamefont {Xiao}},\ }\href {\doibase 10.1103/PhysRevD.90.074026} {\bibfield  {journal} {\bibinfo  {journal} {Phys. Rev. D}\ }\textbf {\bibinfo {volume} {90}},\ \bibinfo {pages} {074026} (\bibinfo {year} {2014})},\ \Eprint {http://arxiv.org/abs/1407.5952} {arXiv:1407.5952 [hep-ph]} \BibitemShut {NoStop}%
\end{thebibliography}%
\bibliographystyle{apsrev4-1}

\end{document}